\documentclass[iop,apj,tighten]{emulateapj} 

\usepackage{amsmath,amstext}
\usepackage[breaklinks,colorlinks,citecolor=blue,linkcolor=magenta]{hyperref}
\usepackage[all]{hypcap} 
\usepackage{mathrsfs}
\usepackage{float}
\usepackage{array}
\usepackage{combelow}
\usepackage[caption=false]{subfig}
\newcommand{\ab}[1]{AS0#1}
\newcommand{\Chandra}{\textit{Chandra}~}

\newcommand{\HST}{\textit{HST}}
\newcommand{\R}{\ensuremath {\mathrm {R_{500}}}}
\newcommand{\M}{\ensuremath {\mathrm {M_{500}}}}
\newcommand{\Msol}{\ensuremath {\mathrm {M_{\odot}}}}

\begin{document}
\title{ \textit{Chandra} observations of the Abell S0295 cluster}
\author{Aurelia \textsc{Pascut}\altaffilmark{1,2}
    and John P. \textsc{Hughes}\altaffilmark{1}
   } 

\altaffiltext{1}{Department of Physics and Astronomy, Rutgers University, 136 Frelinghuysen Road, Piscataway, NJ
     08854-8019, USA; jph@physics.rutgers.edu}
\altaffiltext{2}{\cb{S}tefan cel Mare University, Astronomical Observatory and Faculty of Electrical Engineering and Computer Science, Universit\u{a}\cb{t}ii 13, Suceava, Romania; aureliapascut@gmail.com}

\begin{abstract}
We present deep ($205 \rm ks$), \Chandra observations of the \ab295 binary merging cluster ($z=0.30$). In the X-ray image, the secondary component is clearly visible as a surface brightness peak, while the primary cluster has a flatter distribution. We found cool gas ($\sim 6 \rm keV$) associated with the secondary, while the central temperature of the primary does not deviate significantly from the mean temperature of the cluster of $\sim 9.5 \rm keV$. In the vicinity of the primary's core we found the hottest region in the cluster accompanied by a surface brightness discontinuity. We propose that this region corresponds to a shock, for which we estimate a Mach number of $1.24_{-0.22}^{+0.30}$. We found other merger signatures such as a plume of cool gas emerging from the primary cluster and a cold front and a possible bow shock (Mach number of $1.74_{-0.74}^{+1.02}$) leading the secondary cluster. Based on the observed properties in comparison to binary merger simulations from the literature we propose for 
\ab295 a low mass ratio, off-axis merging scenario, with secondary close to first apocentre. Comparison of our results with strong lensing observations of \ab295 from \cite{Cibirka2018} shows an offset between the total mass peak and the bulk of the gas distribution in the primary cluster. The properties of the merger and the existence of the offset between mass peak and gas make \ab295 a promising candidate for the study of mergers involving non-cool core clusters and the nature of dark matter. 
\end{abstract}

\section{Introduction}

In the context of the standard hierarchical scenario of structure formation, mergers between clusters are key events through which clusters grow \citep{Voit2005,Kravtsov2012}. However, the merging process is not an instantaneous phenomenon and the time it takes for two interacting systems to merge and form a morphologically relaxed, more massive system is typically ${\sim}4-6$ Gyr \citep{Poole2006}. During all this period of time the cluster suffers dramatic changes in its observed properties \citep{Ricker2001,Poole2007}, as a result of the conversion of large amounts of gravitational energy into kinetic energy during the merger event. This unique set-up created during merging, makes cluster mergers sought-after targets for the study of the properties and behavior of the three individual mass components (e.g. dark matter (DM), intracluster medium (ICM) and galaxies) during their self-interaction or interaction with each other.

For example, observations of cluster mergers have been used to confirm the existence of dark matter \citep{Clowe2006}, study its nature and put constraints on the dark-matter self-interaction cross-section \citep{Randall2008,Harvey2015,Wittman2017}.

Compared to the DM and galaxies, the ICM, due to its highly collisional nature, is the cluster mass component which bears the scars of a cluster merger in its thermal and X-ray surface brightness distribution \citep{Sarazin2002,Molnar2015}. 

Therefore, typical observational signatures of merging clusters are contact edges between regions of gas with different entropies \citep{Markevitch2007}. These contact edges can take two forms: cold fronts, which are an increase in surface brightness by a factor of about 2 over a distance of several kpc accompanied by a drop in temperature by a similar magnitude, and shock fronts, in which both, the surface brightness and temperature show a drop across the edge. Moreover, the gas pressure over cold fronts is nearly continuous, while shocks are characterized by discontinuous pressure profiles.

Cold fronts represent the most common feature seen in X-ray observations of merging clusters \citep{Owers2009b,Ghizzardi2010,Botteon2018}, with some clusters displaying more than one cold front \citep{Rossetti2013,Werner2016}. Although cold fronts have been observed predominantly in merging clusters, they also have been found in a significant number of apparently relaxed ones \citep{Markevitch2001,Mazzotta2001,Sanders2005,Dupke2007}. The abundance of cold fronts in merging clusters, may be explained by the  fact that the cold front structure can persist for gigayears and that even a minor merger can displace the low entropy gas and create multiple cold fronts \citep{Ascasibar2006}. 
Numerical simulations of idealized cluster mergers have shown two different mechanisms which can bring two regions of gas with different entropies in contact and form cold fronts: 1) by displacing low entropy gas in regions with higher entropy through sloshing \citep{Tittley2005,Ascasibar2006} and 2) through gas stripping followed by adiabatic cooling \citep{Ricker2001,Nagai2003,Poole2006,ZuHone2011}.

Besides cold fronts, most cluster merging simulations predict the formation of a pair of shocks shortly before the moment of first pericentric passage, shocks which will propagate in opposite directions towards the cluster outskirts. The morphology of the shocks is dependent on the mass ratio of the merging clusters and the geometry of the merger \citep{Paul2011}. The expected Mach number (defined as the ratio between the shock speed and the speed of sound in the pre-shock gas) of these shocks is typically $\le 3$ \citep{Gabici2003,Sarazin2002}. 

Observationally, there are a significant number of disturbed clusters where sudden drops in temperature and/or density have been detected. While most of them correspond to shocks with $\mathscr{M} \sim 1-2$ \citep{Krivonos2003,Markevitch2005,Macario2011,Bourdin2013,Ogrean2014,Ichinohe2015,Eckert2016}, the number of strong shocks ($\mathscr{M} \sim 3$) detected in X-ray observations is limited to a handful of clusters (e.g. Bullet cluster - \cite{Markevitch2002}, Abell 665 - \cite{Dasadia2016}, El Gordo  - \cite{Botteon2016a}, Abell 3667 - \cite{Akamatsu2012,Sarazin2016}). Moreover, clusters where two opposite shocks have been reported based on X-ray observations are extremely limited (Abell 2146 -  \cite{Russell2010, Russell2012},  CIZA J2242.8+5301 - \cite{Ogrean2014,Akamatsu2015}, Bullet cluster - \cite{Shimwell2015}, Abell 2219 - \cite{Canning2017}).

Compared to simulations, observational detection of shocks is constrained by two important factors: 1) merger geometry and 2) stage of the merger. In the first case, small inclination of the merger axis with respect to the plane of the sky leads to projection effects that can wash out the surface brightness discontinuities across a shock, while projected temperature discontinuities decrease more slowly with inclination angle \citep{Akahori2010}. Secondly, the detection of a shock which has propagated to the outer, lower density region of the cluster (where cluster emission is dominated by background emission) requires deep observations for the construction of high resolution, high signal-to-noise ratio of temperature and density profiles. This constraint gets even more difficult to meet when the merger shock is weak.

Cold fronts and shocks in merging clusters represent important tools for the study of cluster dynamics \citep{Markevitch2005,Owers2009,Akamatsu2015,Dasadia2016a,Canning2017}, cluster physics, as well as plasma physics \citep{Vikhlinin2001,Markevitch2006,Russell2012,Datta2014,Chen2017}.  

The target of our study, Abell Supplementary cluster S295 (\ab295) is a massive merging cluster at redshift of 0.30 (see Table \ref{table:global_prop} for cluster properties), discovered optically \citep{Abell1989}. It has subsequently been detected, with high significance, in different cluster surveys based on the SZ effect such as ACT \citep{Menanteau2010,Hasselfield2013}, SPT \citep{Williamson2011} and Planck \citep{PlanckCollaboration2014,PlanckCollaboration2015}. The SZ detection as well as the optical properties of \ab295 have motivated X-ray follow-up of this cluster with \textit{Chandra} and \textit{XMM-Newton} for more detailed studies of this system. A very recent strong-lensing study \citep{Cibirka2018} of this cluster shows the system has a clear bimodal mass distribution which follows the cluster galaxy concentration. 

In this paper we present an X-ray investigation of the merging galaxy cluster \ab295, using 205 ks \Chandra observations. We study the spatial and spectral properties of the intracluster medium, with the aim of understanding the relationship between the observed properties and the dynamical state of the cluster. 

We assume $\rm H_{0}=70~km~s^{-1}~ Mpc^{-1}$, $\rm \Omega_{m}=0.3$ and $\rm \Omega_{\Lambda}=0.7$. Uncertainties are quoted at $68 \%$ confidence level. All coordinates are referenced to the J2000 epoch. All \Chandra X-ray images presented are unbinned images (the pixel size is $0.49 \arcsec$).

 \begin{table}
  \caption{Global properties of \ab295. Redshift and cluster members are from \protect\cite{Ruel2014} and BCG positions for primary ($\rm BCG_E$) and secondary ($\rm BCG_W$) are from \protect\citep{Cibirka2018} and \protect\cite{Zenteno2016}, respectively.}
\label{table:global_prop}

\begin{tabular}{>{\raggedright\arraybackslash}p{3.8cm}%
>{\raggedright\arraybackslash}p{1.9cm}%
>{\raggedright\arraybackslash}p{0.5cm}%
}

\hline
\\
\multicolumn{3}{c}{Optical properties} \\
\\
\cline{1-3}
Redshift  & 0.3001 & \\
Confirmed Members &30  & \\
R.A. $\rm BCG_E$&$02^{\rm h}45^{\rm m}34.97^{\rm s}$ &  \\
Dec $\rm BCG_E$&$-53^{\rm d}02^{\rm m}54.44^{\rm s}$ & \\
R.A. $\rm BCG_W$&$02^{\rm h}45^{\rm m}24.84^{\rm s}$ & \\
Dec $\rm BCG_W$&$-53^{\rm d}01^{\rm m}45.12^{\rm s}$ & \\
\\

\hline
\\
\multicolumn{3}{c}{X-ray derived properties} \\
\\
\cline{1-3}
R.A. centroid &  $02^{\rm h}45^{\rm m}29.66^{\rm s}$ &\\
Dec centroid  &  $-53^{\rm d}02^{\rm m}14.37^{\rm s}$ & \\
\R  &  $1.26 \pm 0.04$  & Mpc \\
\M & $7.6 \pm 0.4$   &  \Msol \\
$\rm kT _{ r<\R}$&  $9.51 \pm 0.32 $ &  keV\\
$\rm kT _{ 0.15<r<\R}$ & $9.17 \pm 0.48$ & keV\\
Centroid shift &   0.05 \R \hspace{-1.5cm} & \\
\end{tabular}
 \end{table}

\section{Data reduction}
\label{sec:data_reduction}

This study uses all available \Chandra observations of the \ab295 cluster (6 observations with ObsIDs: 12260, 16127, 16524, 16525, 16526, 16282). The cluster was observed for a total exposure of 205.7 ks.

All observations were taken using the Advanced CCD Imaging Spectrometer (ACIS; \cite{Garmire1992,Bautz1998}), with the ACIS-I array at the focal plane and in very faint (VFAINT) telemetry format. The VFAINT mode allows a better screening of particle background events, therefore reducing significantly the level of particle background. 
Data reduction was performed using CIAO (version 4.7), \textit{Chandra}'s data analysis system, and calibration files from the \Chandra Calibration DataBase (CALDB; version 4.6.7). All observations have been reprocessed starting from level1 event files using the CIAO \textbf{chandra\_repro} script which creates an observation specific bad pixel file and applies corrections using the latest calibration files. It also applies the VFAINT mode filtering of particle background events and includes only events with grades = 0,2,3,4,6 and status=0  in the final filtered event file used for data analysis. 

The cleaning of background for potential anomalous periods of very high background rates (flares)  has been done by applying a 3 sigma clipping algorithm to the light curves extracted from the entire field of view, excluding bright sources, in the energy band of 0.5-12 keV and binned in a time interval of 250 s. We found that none of the observations was affected by serious flares and the final exposure time, after flare filtering is 205 ks.

\section{Data analysis}
\label{sec:data_analysis}

\subsection{Image analysis}
\label{subsec:sb}

\begin{figure*}\centering
 \includegraphics[width=0.95\textwidth,keepaspectratio,trim={1.0cm 0.7cm 0cm 4.0cm}, clip=true]{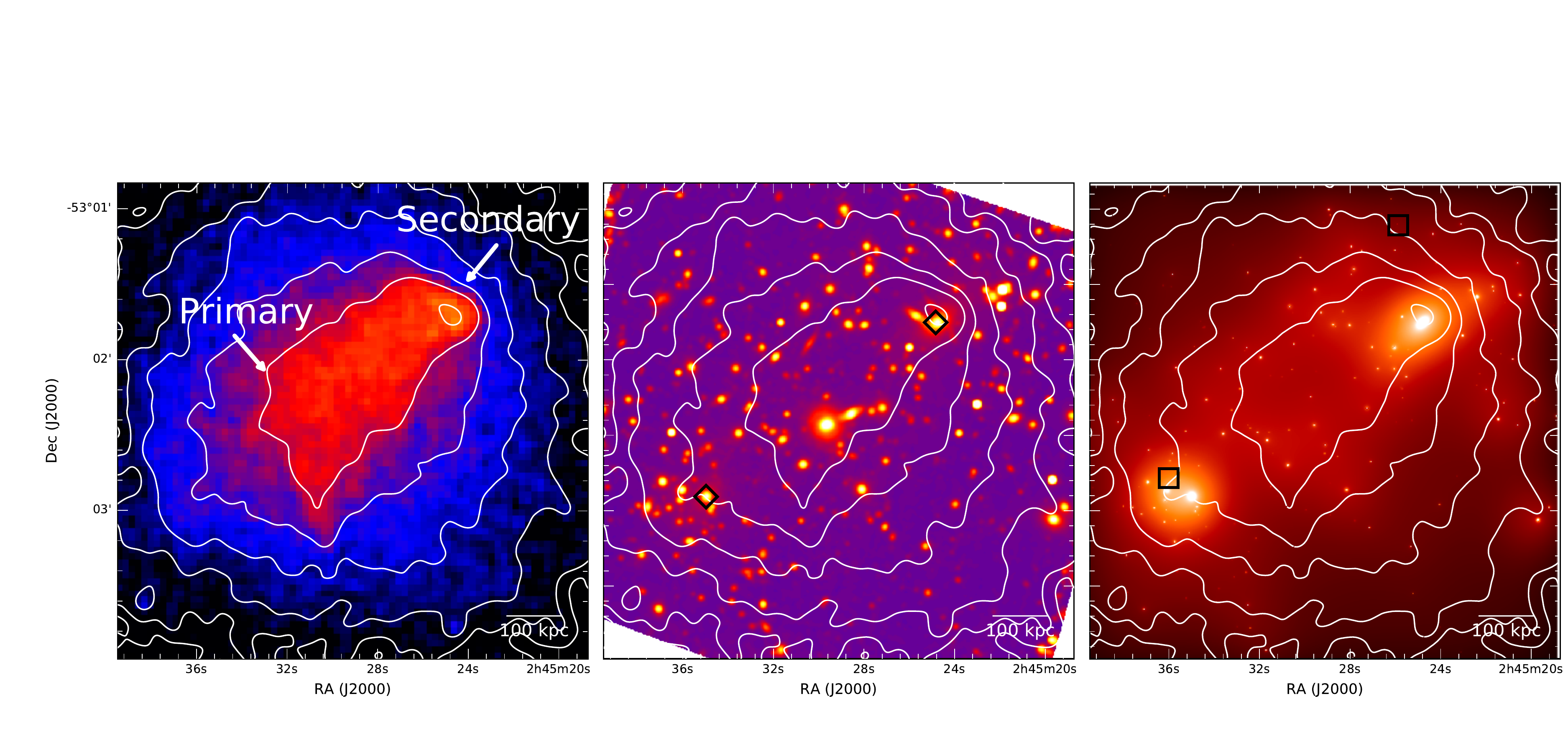}
 \caption{Left: \Chandra soft band (0.5-2.0 keV), exposure corrected, smoothed image of \ab295 cluster. The secondary cluster (located in the NW part of the image) is clearly visible as a surface brightness peak while the primary cluster (in the SE) has a flatter surface brightness distribution. Center: \HST\ archival image taken in the ACS/WFC instrument/detector configuration (PI: F. Pacaud). Right: Projected surface mass-density distribution obtained from strong lensing analysis of \ab295 by \protect\cite{Cibirka2018}. All images are matched for the same coordinates. The black diamond symbols (center panel) and square symbols (right panel) mark the positions of the brightest cluster galaxies (see also Table \ref{table:global_prop}) and the approximate positions of two radio relics detected by \protect\cite{Zheng2018}.}
\label{fig:AS0295_image}
\end{figure*}

The X-ray image obtained after applying the data reduction steps described in Section \ref{sec:data_reduction} and merging all individual observations is presented in the left panel of Figure  \ref{fig:AS0295_image}. The merged image of all six individual observations was obtained from a merged event file, created by reprojecting each event file to the same tangent point, which is the tangent point of the observation with the longest exposure time. No absolute astrometry corrections were applied to individual observations before merging. The image in Figure  \ref{fig:AS0295_image} is a soft band (0.5-2.0 keV), background included, exposure corrected image, smoothed with a $1.5 \arcsec$ Gaussian kernel.

The central panel of Figure \ref{fig:AS0295_image} shows a 2 ks optical image taken with the {\textit Hubble Space Telescope} (\HST), using the Advanced Camera for Surveys (ACS) instrument and Wide Field Channel (WFC) detector. The level of image processing is the one applied as part of the \HST\ calibration pipeline, and includes: bias and dark current subtraction and flat-fielding. Since during this process no astrometric corrections have been applied, the positional uncertainty is $ \sim 1-2 \arcsec $. With an uncertainty for \Chandra data of $\le 1\arcsec$, the combined \Chandra and \HST\ positional accuracy is within 2.2 $\arcsec$ (which is 10 kpc at the cluster's redshift of 0.3).

The right panel presents the projected surface mass-density distribution obtained from the strong lensing analysis of \ab295 by \cite{Cibirka2018}.

\begin{figure}[h]\centering
\includegraphics[width=0.5\textwidth,keepaspectratio]{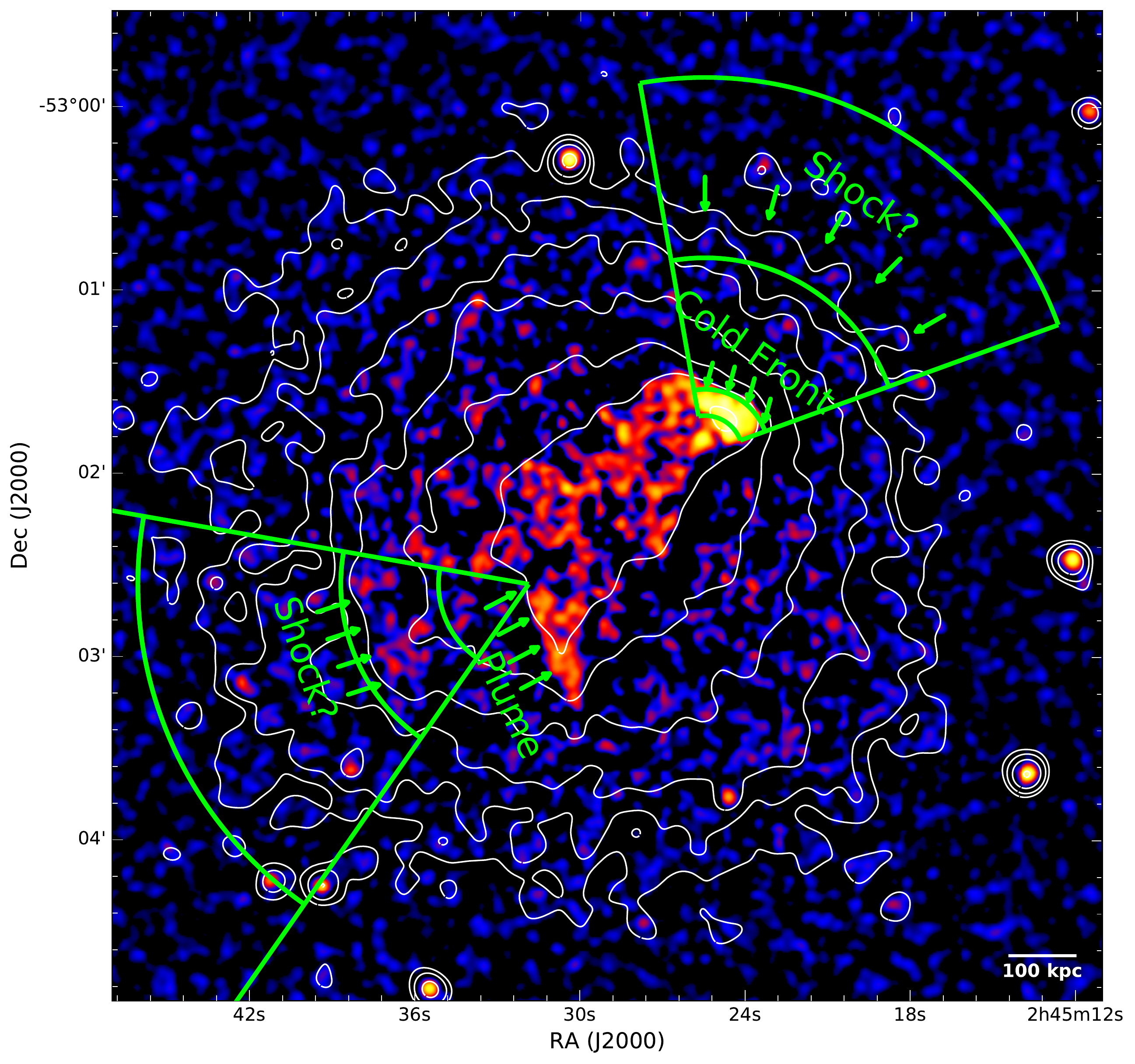}
 \caption{Unsharp-masked image created by subtracting a $\sigma = 9.8 \arcsec$ Gaussian smoothed image of the cluster from a $\sigma=1.5 \arcsec$ Gaussian smoothed image. Green regions are used for the creation of surface brightness and temperature profiles shown in Figures \ref{fig:secondary_CF_sb_kt} and \ref{fig:primary_shock_sb_kt}}
\label{fig:unsharp-masked}
\end{figure}

To enhance the existing X-ray surface brightness features, we have created an unsharp-masked image by subtracting a highly smoothed image (using a Gaussian kernel of size $\sigma = 9.8 \arcsec$) from the image shown in Figure \ref{fig:AS0295_image}, which has been smoothed with a $1.5 \arcsec$ Gaussian.  The unsharp-masked image is shown in Figure \ref{fig:unsharp-masked}.

\subsection{Spectral analysis}
\label{subsec:tempmap}

To investigate the variation of gas temperature across the cluster, a 2D map of projected gas temperature was created (Figure \ref{fig:tempmap}). 

To create the map, the full band (0.5-7.0 keV) image of the cluster is divided into multiple regions, using the adaptive binning algorithm of  \cite{Diehl2006}. This algorithm assigns each pixel in the image to a region such as to obtain in each region a chosen target signal-to-noise ratio. We choose a signal-to-noise ratio of 40 for the binning, a value high enough to allow for sufficient numbers of counts per bin for accurate spectral fitting. For the adopted signal-to-noise, the adaptive binning algorithm results in 38 regions with the net number of counts varying between 1300 and 3900 among the regions, and a mean of $\sim 2000$ counts.

To estimate the temperature for each region, spectra, together with corresponding response files are extracted for each region from each individual observation. Similarly, background spectra are extracted from an annular region centered on the cluster and with inner radius of 1.5 Mpc. The gas temperature is estimated by simultaneously fitting the background subtracted spectra with an absorbed  thermal plasma model (XSWABS*XSAPEC model in CIAO) in which the redshift and abundance parameters for thermal component are both frozen to a value of 0.3 and the absorption  is frozen to the Galactic value of $\rm 3.18 \times 10^{20} ~cm^{-2}$ \citep{Dickey1990}. The 0.3 value chosen for the abundance represents the typical metallicity observed in low redshift clusters (\cite{Mushotzky1997},\cite{DeGrandi2001},\cite{Snowden2008}).
 The same spectral fitting method is adopted for creating the temperature map and temperature profiles which will be used in the following sections to investigate temperature variations across several interesting regions in the cluster.

\section{Global properties}
\label{sec:global_properties}

Evidence for the presence of an on-going merger in \ab295 comes from the irregular morphology of the X-ray gas as seen in the X-ray image of the cluster and the shape of the X-ray contours (see Figure \ref{fig:AS0295_image}, left panel). The secondary cluster is clearly visible as an X-ray peak in the NW direction. The orientation of the merger axis in the SE-NW direction is suggested by the elongated morphology of the X-ray emission along this direction and the ``bunching up'' of contour lines in front of the secondary. This accords well with the SE-NW orientation of the two total mass peaks visible in the strong lensing map (Figure \ref{fig:AS0295_image}, right panel), as well as the optical galaxy distribution (Figure \ref{fig:AS0295_image}, central panel).

Several X-ray global properties of the cluster derived from our data analysis as presented below, together with a few optical properties taken from the literature are summarized in Table \ref{table:global_prop}. 

The position of the centroid is estimated from the smoothed, exposure corrected image, iteratively, using a circular region with a radius of $3 ~ \rm arcminutes$. At the first step, the center of the circular region is the peak of the cluster and a first estimate of the centroid is obtained. In the following steps, the position of the centroid is reestimated using the new value for the centroid as the center for the circular region and the process is repeated until convergence.

To quantify the degree of disturbance of the ICM, we have calculated the centroid shift parameter, following the method described in \cite{Poole2006}. Centroid shift is the standard deviation of distances between the X-ray peak and centroid measured within several circular regions with radii decreasing from \R\ to $0.01$\R, with a step of $0.05$\R. For each new radius, the circular region is centered on the X-ray peak, while the centroid is reestimated and therefore a new separation between peak and centroid is calculated. Following \cite{Poole2006}, when estimating the centroid, we excluded a circular region of 30 kpc radius, centered on the peak position.

In general, clusters are characterized by centroid shifts (as a fraction of \R, the radius enclosing 500 times the critical density of the Universe) ranging from $\sim 0.001$ to $\sim 0.15$ and a value of 0.01-0.02 has been adopted empirically as a threshold between relaxed and disturbed clusters \citep{OHara2006,Maughan2008a,Bohringer2010,Cassano2010,Weissmann2013,Donahue2016}. With a centroid shift of 0.05, \ab295 can be clearly classified as a disturbed cluster.

\R\ has been estimated using the $\rm M-Y_X$ relation from \cite{Vikhlinin2009}, where $\rm Y_X$ is the product between gas temperature and mass. Following their iterative prescription, we have estimated at each step the gas temperature from a spectra extracted within 0.15-1\R, and the gas mass by fitting the emissivity profile obtained from the same region with a gas density model \citep{Vikhlinin2009} projected along the line of sight. In the first step, the value assigned to \R\ is $0.2 ~\rm Mpc$ and these steps are repeated until convergence of \R\ is obtained. As in \cite{Vikhlinin2009}, at all but the first step we have excluded the central 0.15\R\ from temperature estimation to avoid contamination from a possible cool core, which are known to introduce scatter in scaling relations. Since the secondary has cool gas associated with its core, we have treated the secondary as a contaminating source and exclude a circular region, centered on the secondary's X-ray peak and with a radius of $\sim 80 \rm kpc$ 
(large enough to include the bulk of X-ray emission) when estimating \R. We obtain very similar results for \R\ when we treat the secondary as a contaminating source compared to the case when secondary is not excluded from the analysis. 

Results from Table \ref{table:global_prop} show that \ab259 is a morphologically disturbed (centroid shift of 0.05), hot ($\sim \rm 9~ keV$), massive ($\sim 8\times10^{14}~\Msol$) cluster. 


\section{Different types of surface brightness features}
\label{sec:as0295}

Since mergers between galaxy clusters leave their imprint on the ICM in the form of shocks, cold fronts and other transient features observed in the gas surface brightness and temperature distribution, we next look for these kind of features for \ab295 in our X-ray image and temperature map. 

A qualitative analysis of the unsharp-masked image (see Section \ref{subsec:sb} and Figure \ref{fig:unsharp-masked}) shows two clear features in the form of surface brightness edges (one near the primary, labeled ``plume'' and the other in the secondary labeled ``cold front'' in Figure \ref{fig:unsharp-masked}). The most striking feature is the edge seen in the NW direction, very close to the core of the secondary cluster. Another noticeable feature, which is also apparent in the X-ray image and contours (Figure \ref{fig:AS0295_image}) is the sharp edge seen close to the core of the main cluster, in the S direction. Two other  less obvious features labeled ``Shock?'' in Figure \ref{fig:unsharp-masked} are the locations of possible shocks that are studied in detail below.

\begin{figure}[h]\centering
\includegraphics[width=0.45\textwidth,keepaspectratio,trim={3cm 7.5cm 4.5cm 0.5cm}, clip=true]{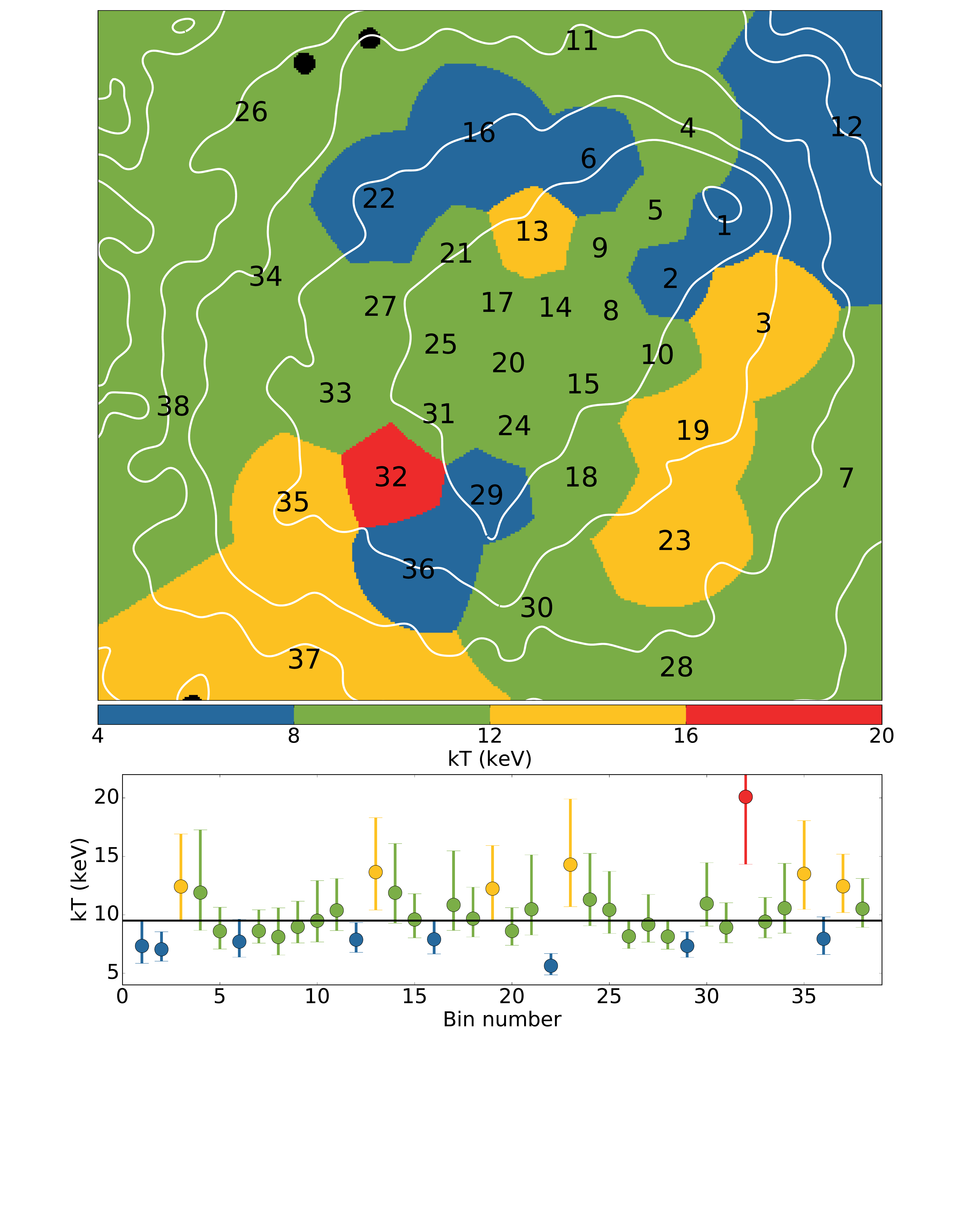}
 \caption{Top panel: Temperature map created using \textit{Chandra} data over the 0.5-7.0 keV band. The numbered polygonal regions in the figure were generated by an adaptive binning algorithm \protect\citep{Diehl2006} so as to enclose a target signal-to-noise ratio of 40. Bottom panel: Estimated projected temperature for each region shown in the temperature map. The horizontal line at 9.5 keV marks the global mean temperature of the cluster estimated within a circular region with radius equal to \R.}
\label{fig:tempmap}
\end{figure}

Before investigating in detail the nature of these features seen in surface brightness, we look at the temperature distribution in the cluster as shown in Figure \ref{fig:tempmap}. The first thing to notice is the presence of a cold region situated in the NW direction (region 2 in the temperature map) and which coincides with the subcluster. The high temperature difference seen in the temperature map for this region compared to an adjacent region (region 5) and the high contrast in surface brightness seen in Figure \ref{fig:unsharp-masked} suggests this might be  a cold front. 

Related to the other two features seen in surface brightness, the temperature map does not give more information since the size of the bins used to extract temperatures are larger than the size of the edges seen in the surface brightness images. However, we note that the edge seen to the SE of the primary is in the vicinity of the hottest region found in the temperature map (region 32).

In the following sections we investigate the nature of all four surface brightness features identified in the unsharp-masked image.

\subsection{Cold front in secondary cluster}
\label{subsec:CF}
\begin{figure}[h]\centering
\includegraphics[width=0.50\textwidth,keepaspectratio,angle=-90,trim={1.8cm 0.7cm 0cm 0cm}, clip=true]{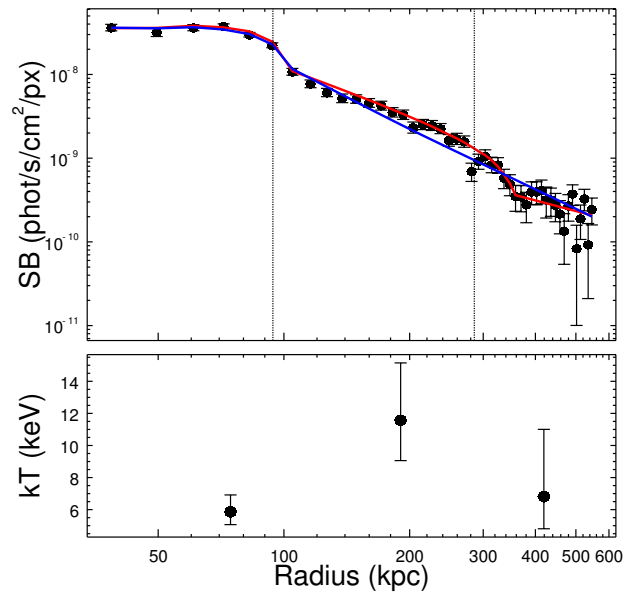}
 \caption{Surface brightness (top panel) and temperature (bottom panel) profiles extracted within annular sectors across the surface brightness drop seen in front of secondary cluster. Vertical lines mark the radii of annular regions used to create the temperature profile. Blue and red lines represent the fit of a density model represented by two and three power laws, respectively, connected by density jumps.}
\label{fig:secondary_CF_sb_kt}
\end{figure}

The most striking feature seen in the unsharp-masked image of the cluster (Figure \ref{fig:unsharp-masked}) is the sharp drop in surface brightness in front of the secondary cluster. More quantitatively, this sudden change in surface brightness is seen as a discontinuity at $\sim 100$ kpc in the surface brightness profile which is shown in the top panel of Figure \ref{fig:secondary_CF_sb_kt}. The profile was extracted  within the spherical annular sector shown with a solid green line in Figure \ref{fig:unsharp-masked} and the three inscribed sectors are used to extract the temperature profile, presented in the bottom panel of Figure \ref{fig:secondary_CF_sb_kt}.

We fit first the surface brightness profile with a model that assumes a density distribution that follows a broken power law, a model commonly used to represent the observed jumps in the distribution of surface brightness. In our case, such a model represents a poor match to the data (blue curve in Figure \ref{fig:secondary_CF_sb_kt}), especially between $200-300 \rm kpc$ where it shows some surface brightness excess above the power law.  Adding another break in the model of density distribution increases significantly the quality of the fit (based on an F-test with with p-value $=0.046$). 

From the fit results we find a jump in density of $2.19 \pm0.29$ at $104$ kpc. A jump with a similar amplitude is also found for the projected gas temperature. The gas temperature increases by a factor of $1.97_{-0.51}^{+0.69}$, from $5.89_{-0.83}^{+1.02}$ keV in the region interior to the discontinuity to $11.59_{-2.54}^{+3.55}$ keV in the region situated  just beyond the discontinuity (see bottom panel of Figure \ref{fig:secondary_CF_sb_kt}). This sudden decrease in surface brightness accompanied by an increase in temperature, with a similar amplitude, is an indication of the presence of a cold front. 

For completeness, we mention that for the second surface brightness discontinuity, situated at a distance of 246 kpc from the cold front, the amplitude of the density jump obtained from the fit is $2.13 \pm 0.87$. The nature of this feature is investigated in Section \ref{subsec:shocks}.

\subsection{Plume in primary cluster}
\label{subsec:plume}
Apart from the cold front detected in front of the secondary cluster, another significant feature seen in the unsharp-masked image (Figure \ref{fig:unsharp-masked}) is the sharp surface brightness edge found to the south of the primary cluster's core. This feature is also clearly visible in the surface brightness contours. 

What is remarkable about this structure, whose length is roughly 240 kpc, is the linear-like shape of its edge. We note that this feature is not an artifact due to a chip gap, since in all observations this region of the cluster falls entirely on a single chip. 

\begin{figure*}\centering
\captionsetup[subfigure]{labelformat=empty}
\centering
\subfloat[]{\includegraphics[width=0.4\textwidth,keepaspectratio,angle=-90,trim={2cm 0.5cm 1cm 5cm}, clip=true]{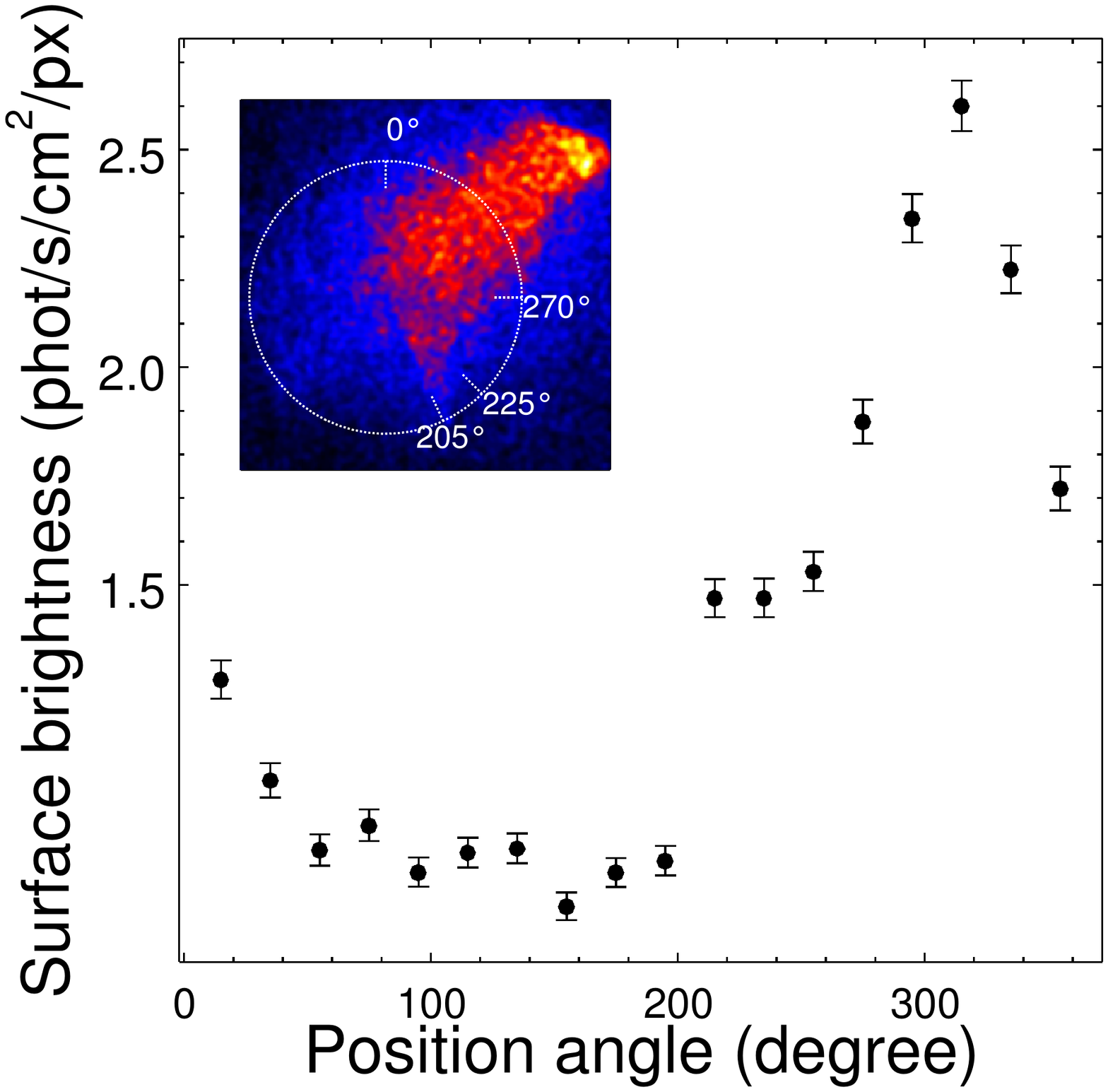}}
\subfloat[]{\includegraphics[width=0.4\textwidth,keepaspectratio,angle=-90,trim={2cm 0.5cm 1cm 5cm}, clip=true]{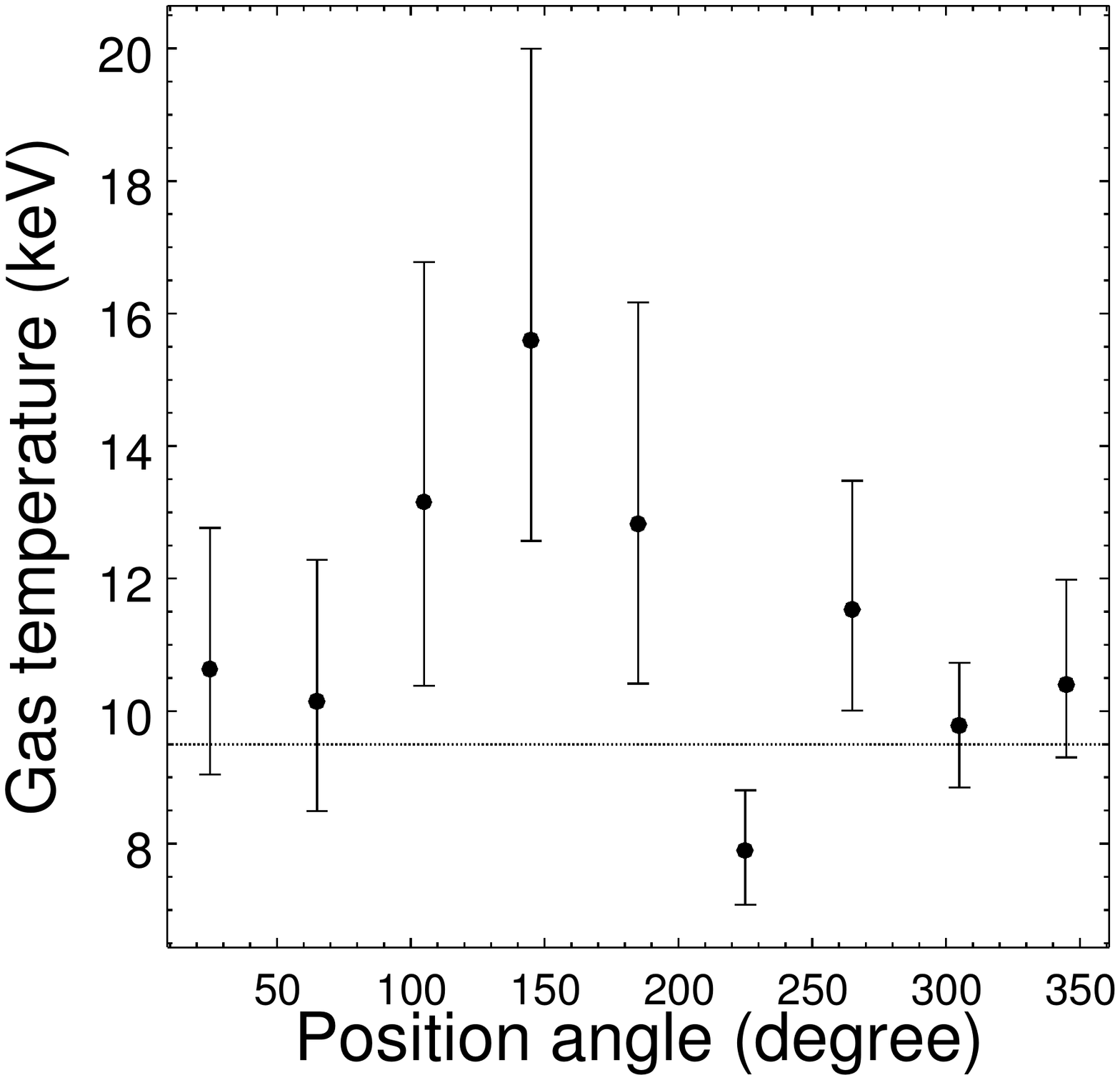}}
 \caption{Left: Azimuthal surface brightness profile extracted using sectors of a circular region centered on the primary cluster. This circular region, overlaid on the X-ray image of the cluster is shown in the inset.  The edge in surface brightness is seen at around 210 degrees. Right: Temperature profile created using spectra extracted from regions similar to those used for surface brightness profile but with twice the opening angle for sectors. The dotted horizontal line marks the global temperature of the cluster of $\rm 9.5 ~ keV$. }
\label{fig:plume}
\end{figure*}

The left panel of Figure \ref{fig:plume}, represents an azimuthal surface brightness profile obtained using sectors centered on the main cluster and with opening angles of 20 degrees. The center of all sectors has been selected such that the edge seen in the unsharp-masked image marks the side of one of the sectors (this sector is marked in the inset in Figure \ref{fig:plume} at angles between 205 and 225 degrees). As already obvious from the cluster X-ray image (the inset in the Figure), the profile shows that the highest surface brightness is in the direction joining the two subclusters (265 - 5 degrees). The surface brightness drops rapidly with increasing position angle and in the opposite direction (85 - 205 degrees) it drops by a factor of $2.53 \pm 0.05$. A sudden jump in surface brightness is seen for the 205-225 degree sector, whose side matches the surface brightness edge.

The right panel of Figure  \ref{fig:plume} shows the temperature profile obtained by fitting spectra extracted from similar regions to those used for surface brightness profile, but with twice the size of the sector's opening angle. It can be seen that the region delineated by the surface brightness edge (205-245 degree) has the lowest temperature of $7.90_{-0.82}^{+0.90}$ keV. Compared with the mean temperature of the two adjacent sectors, the temperature in this region drops by a factor of $1.5_{-0.2}^{+0.3}$. The almost constant temperature of regions enclosed between 270-90 degrees, which include the gas in the direction joining the two clusters, matches very well the mean temperature of the cluster of $\rm 9.5 ~keV$.  The three regions with temperatures above cluster's global temperature correspond to the SE region of the cluster. The temperatures of these three regions are very consistent with the results from the temperature map (Figure \ref{fig:tempmap}), which shows that the highest temperature in 
the 
cluster corresponds to this SE region (region number 32 in the temperature map). 

Based on the presented results we report the detection of a narrow structure of cool gas. The possible origin of this plume of gas will be discussed in more detail in Section \ref{subsec:comparison_sim}.

\subsection{SE discontinuity}
\label{subsec:SE_discontinuity}

Parallel to the plume extending to the south of the primary cluster there is another surface brightness discontinuity visible in the unsharp-masked image (Figure \ref{fig:unsharp-masked}). Although this feature is the weakest one compared to the plume found in the vicinity of the primary and the cold front in front of secondary, its existence and most importantly its position close to the hottest region of the cluster (region 32 in Figure \ref{fig:tempmap}) led us to investigate in detail its properties and the possible nature of this feature.

\begin{figure}[h]\centering
\includegraphics[width=0.48\textwidth,keepaspectratio,angle=-90,trim={1.8cm 1.3cm 1.5cm 0.0cm}, clip=true]{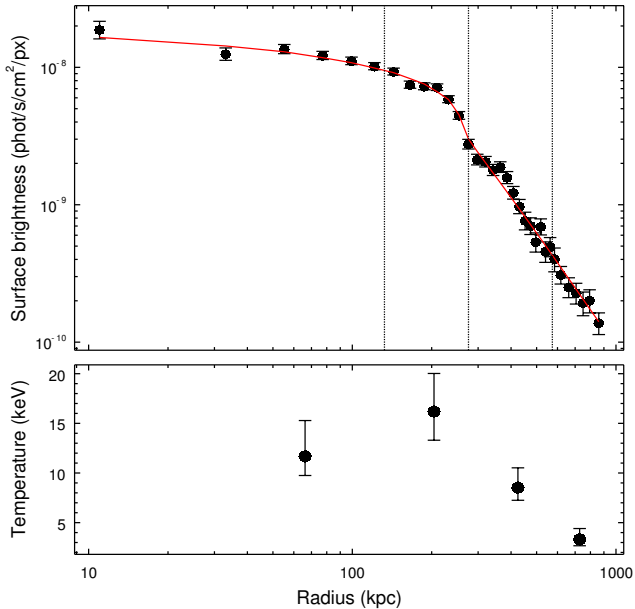}
 \caption{Surface brightness (top panel) and temperature (bottom panel) profile estimated from a sector across the edge found in unsharp-masked image, in the SE part of primary cluster. The red curve represents the result obtained by fitting the surface brightness profile with a line-of-sight integrated density model composed of a broken power law. }
\label{fig:primary_shock_sb_kt}
\end{figure}

Figure \ref{fig:primary_shock_sb_kt} shows the surface brightness (top panel) and temperature profile (bottom panel) in the sector shown in Figure  \ref{fig:unsharp-masked}. 

A sudden change in the distribution of surface brightness is seen at about 250 kpc. Assuming an underlying density profile which follows a broken power law and spherical symmetry, we fitted the projected density along the line of sight to the observed surface brightness. The best fit is shown as a red curve in the upper panel. The fit corresponds to a model in which the density jump has an amplitude of $1.35\pm0.03$. Similarly, comparing the projected temperature found in the immediate vicinity of the discontinuity, we find a temperature jump by a factor of $1.89_{-0.45}^{+0.61}$. The detection of a sudden increase in density accompanied by a rise in temperature is indicative of a shock.

If we explain the observed density discontinuity by the presence of a shock, then the shock's Mach number, estimated from the amplitude of the density jump, is $ M =1.24 \pm0.02$.

An independent estimation of $M$ can be done using the jump in temperature across the shock. Unfortunately, our data do not allow the extraction of a high resolution deprojected temperature profile to estimate the jump in temperature. Using the jump of  $1.89_{-0.45}^{+0.61}$ in projected temperature we estimate a $M=1.7_{-0.4}^{+0.6}$.

\section{Discussions}
\label{sec:discussions}
\subsection{Comparison with simulations}
\label{subsec:comparison_sim}
To learn more about the dynamical state of \ab295, we compare our observations with the results of hydrodynamical simulations of binary galaxy cluster mergers available in the literature \citep{Poole2006,Ascasibar2006,ZuHone2011}. These simulations have shown that different sets of merger parameters such as cluster mass ratio, inclination angle, impact parameter, stage of merging process lead to unique morphologies in the surface brightness and temperature distribution. In some cases, these features are stable against significant variations in the values of dynamical parameters.

\begin{figure*}\centering
\captionsetup[subfigure]{labelformat=empty}
\centering
\subfloat[]{\includegraphics[width=0.23\textwidth,keepaspectratio,trim={0cm 0cm 0cm 0.0cm}, clip=true]{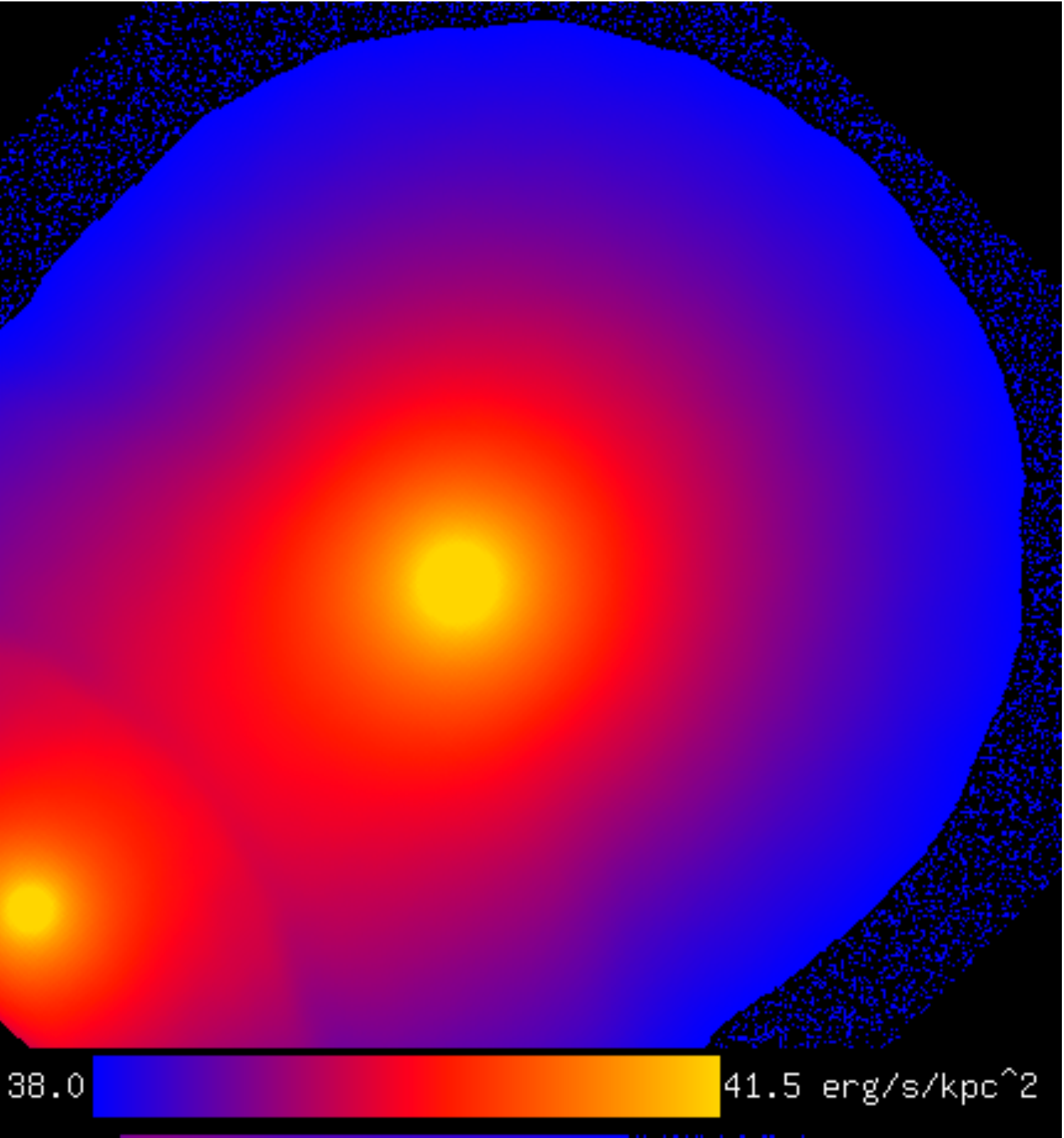}}\hspace{0.05cm}
\subfloat[]{\includegraphics[width=0.23\textwidth,keepaspectratio,trim={0cm 0cm 0cm 0.0cm}, clip=true]{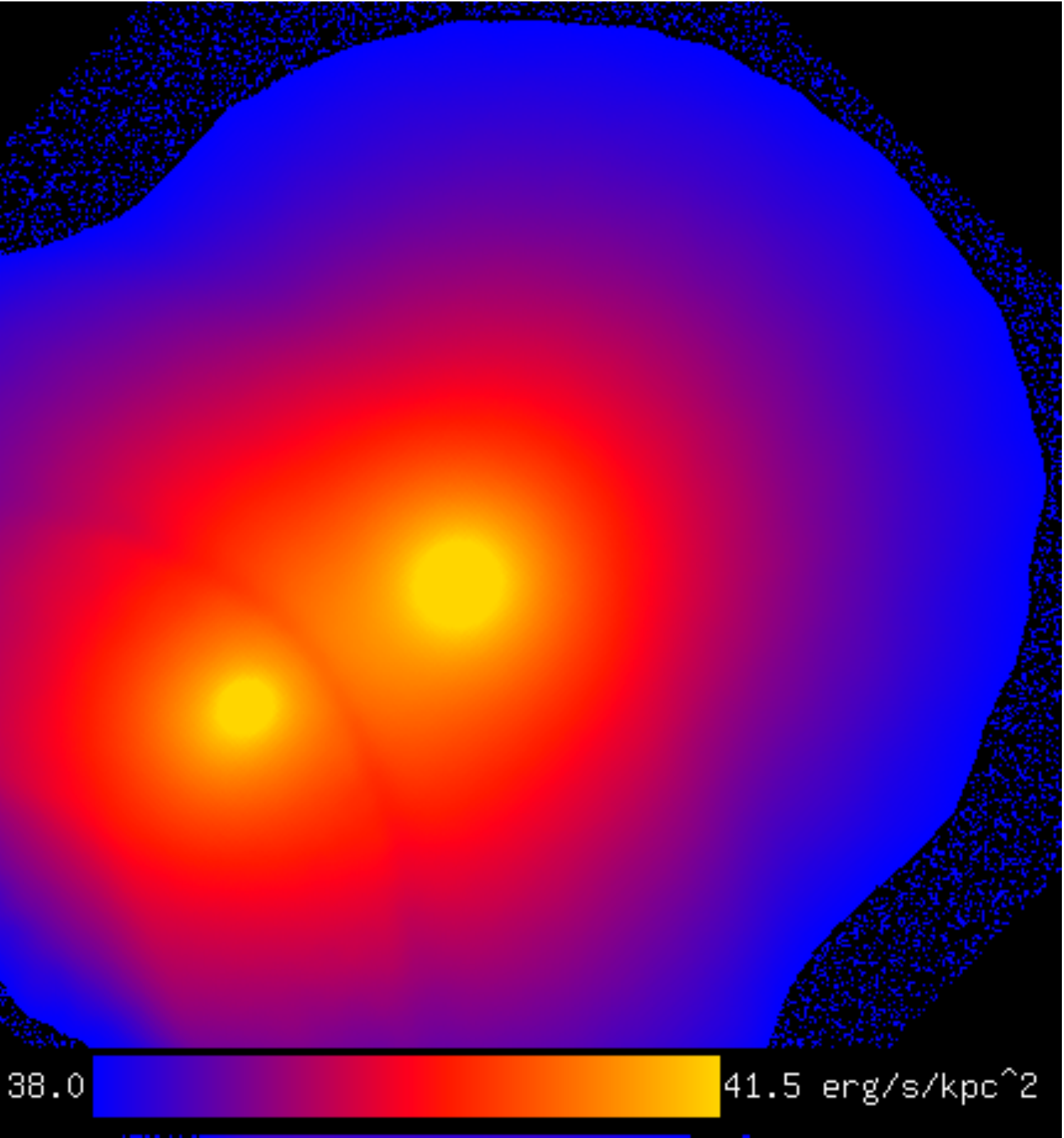}}\hspace{0.05cm}
\subfloat[]{\includegraphics[width=0.23\textwidth,keepaspectratio,trim={0cm 0cm 0cm 0.0cm}, clip=true]{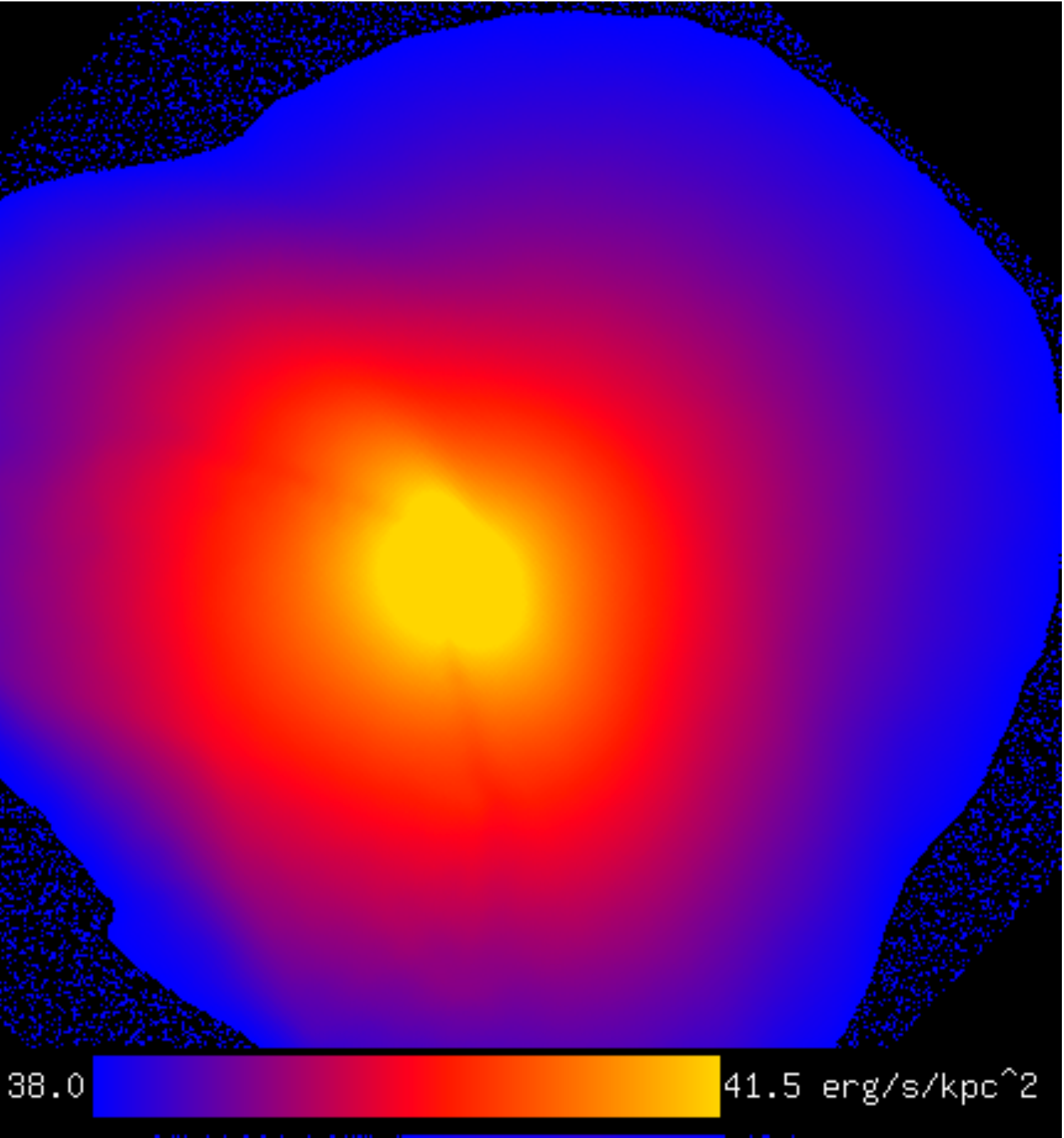}}\hspace{0.05cm}
\subfloat[]{\includegraphics[width=0.23\textwidth,keepaspectratio,trim={0cm 0cm 0cm 0.0cm}, clip=true]{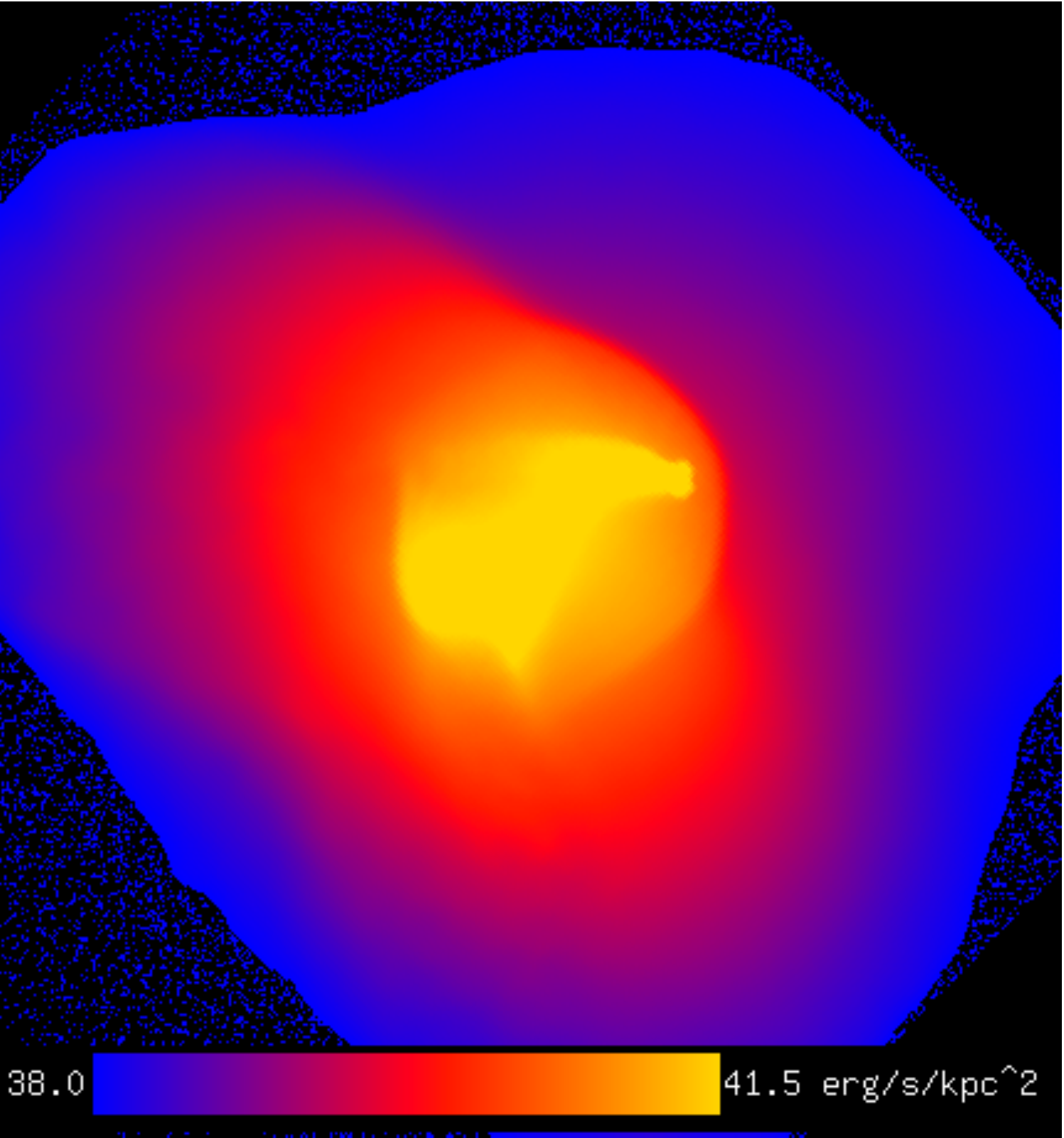}}\vspace{-0.5cm}
\\
\subfloat[]{\includegraphics[width=0.465\textwidth,keepaspectratio,trim={0cm 0cm 0cm 0.0cm}, clip=true]{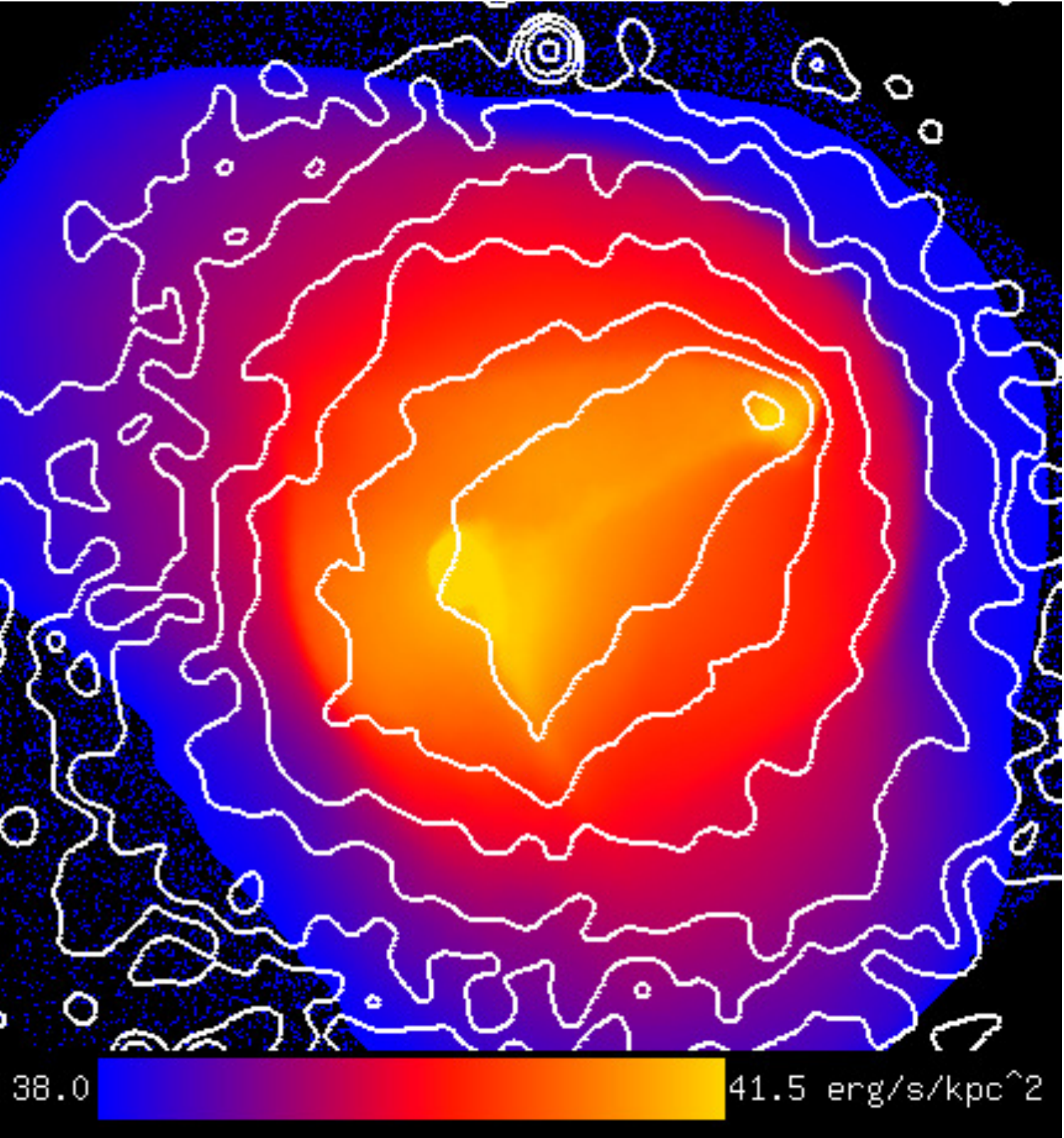}}\hspace{0.2cm}
\subfloat[]{\includegraphics[width=0.465\textwidth,keepaspectratio,trim={0cm 0cm 0cm 0.0cm}, clip=true]{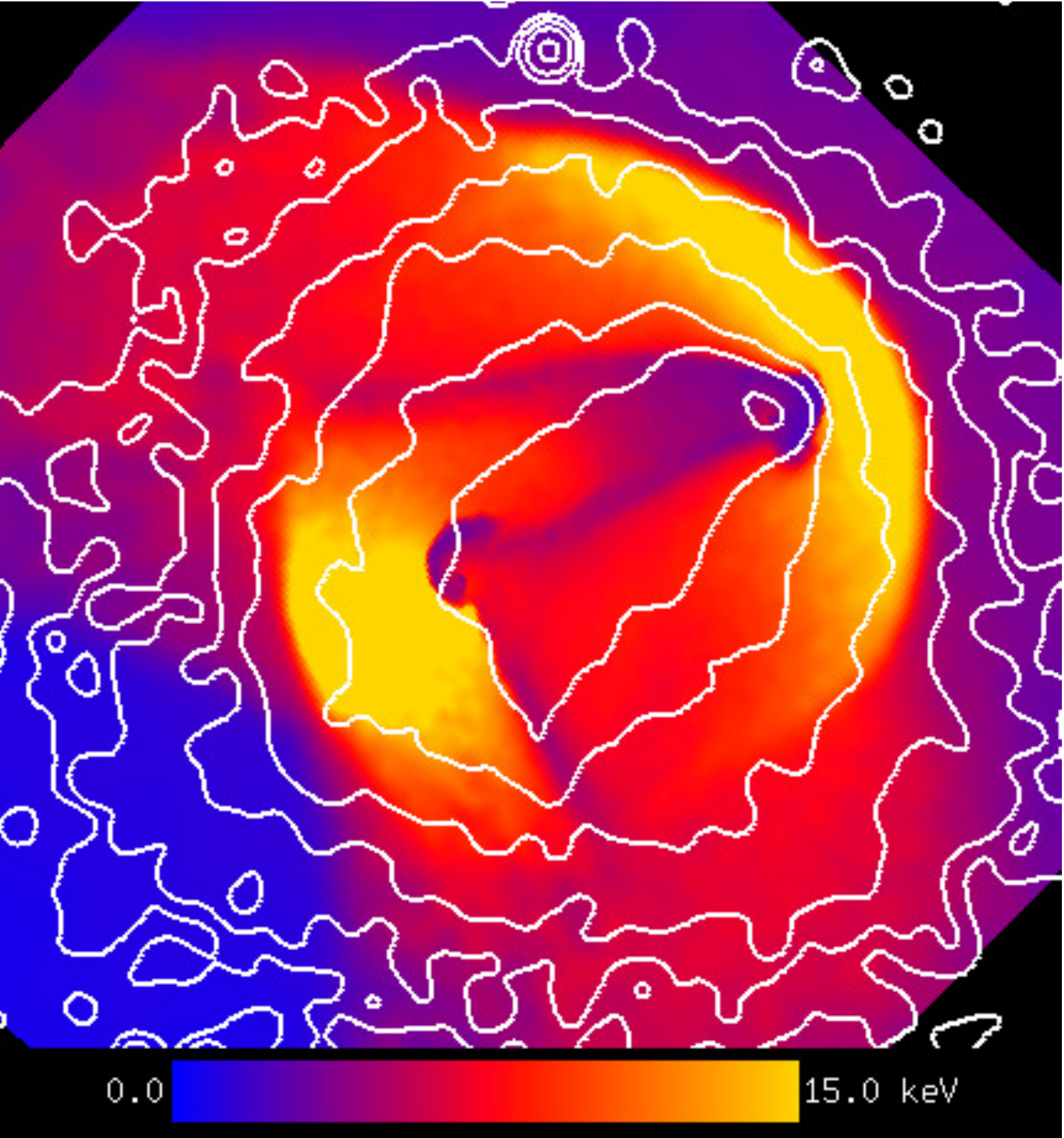}}\vspace{-0.5cm}
\\
\subfloat[]{\includegraphics[width=0.23\textwidth,keepaspectratio,trim={0cm 0cm 0cm 0.0cm}, clip=true]{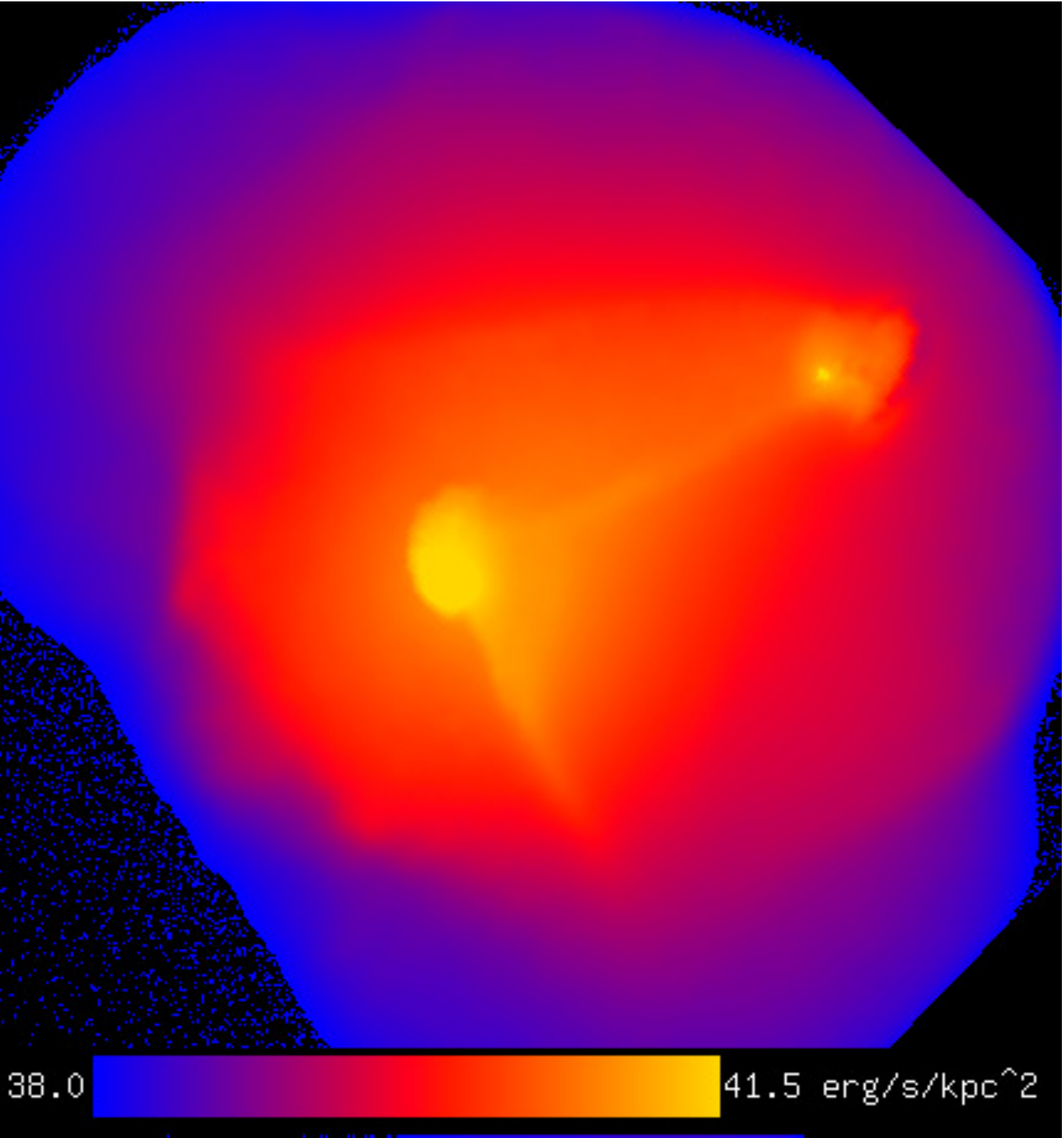}}\hspace{0.05cm}
\subfloat[]{\includegraphics[width=0.23\textwidth,keepaspectratio,trim={0cm 0cm 0cm 0.0cm}, clip=true]{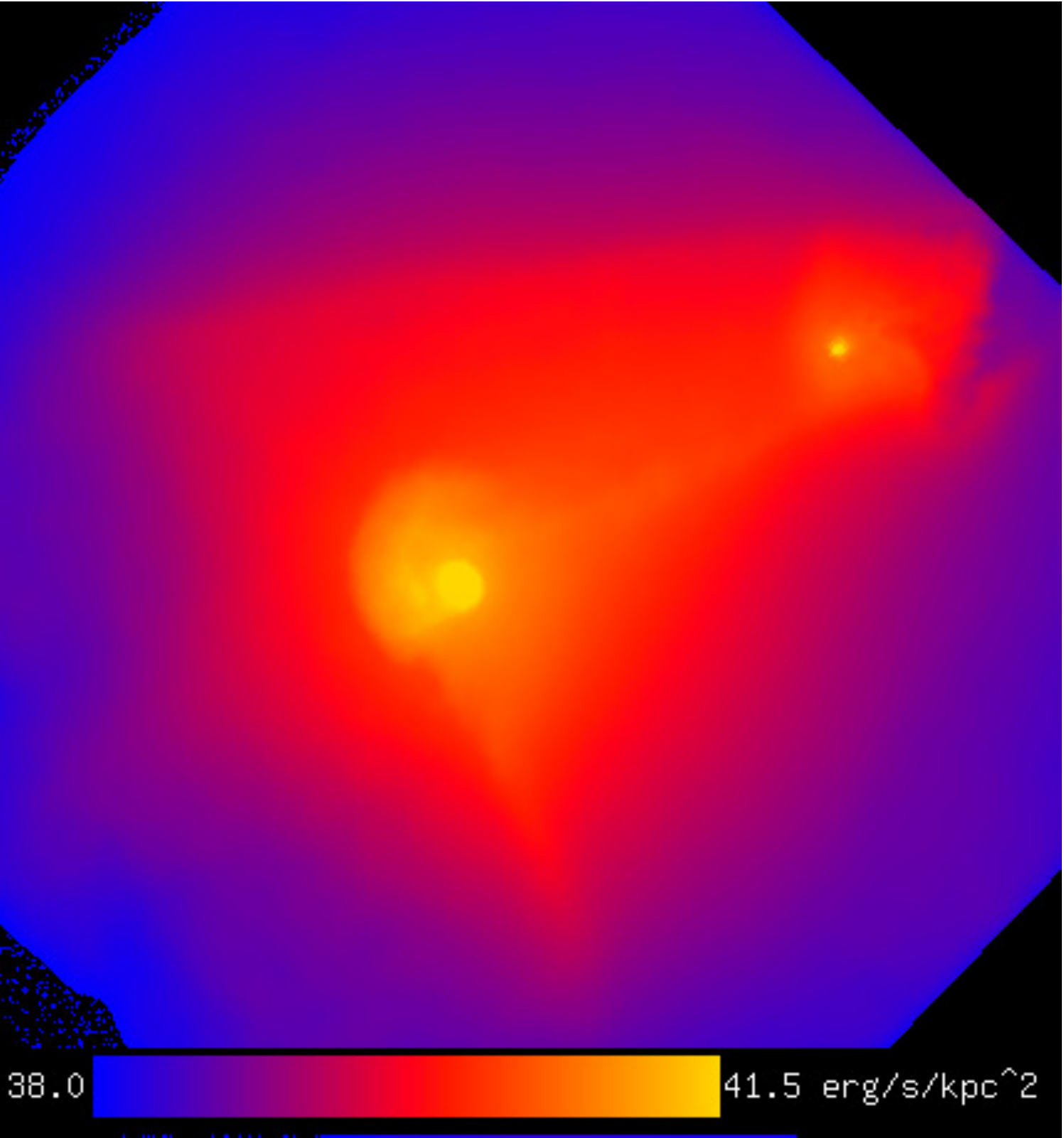}}\hspace{0.05cm}
\subfloat[]{\includegraphics[width=0.23\textwidth,keepaspectratio,trim={0cm 0cm 0cm 0.0cm}, clip=true]{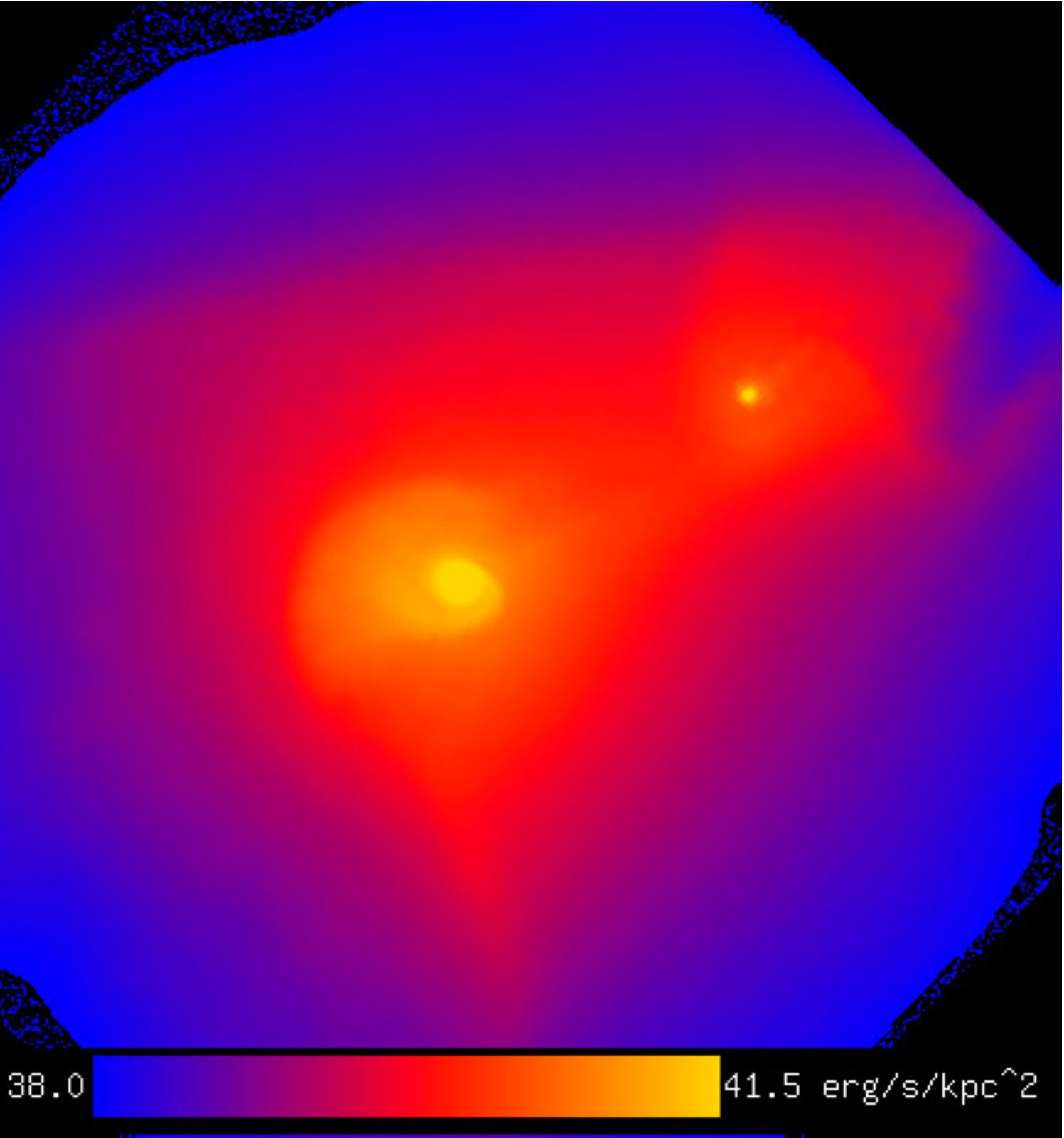}}\hspace{0.05cm}
\subfloat[]{\includegraphics[width=0.23\textwidth,keepaspectratio,trim={0cm 0cm 0cm 0.0cm}, clip=true]{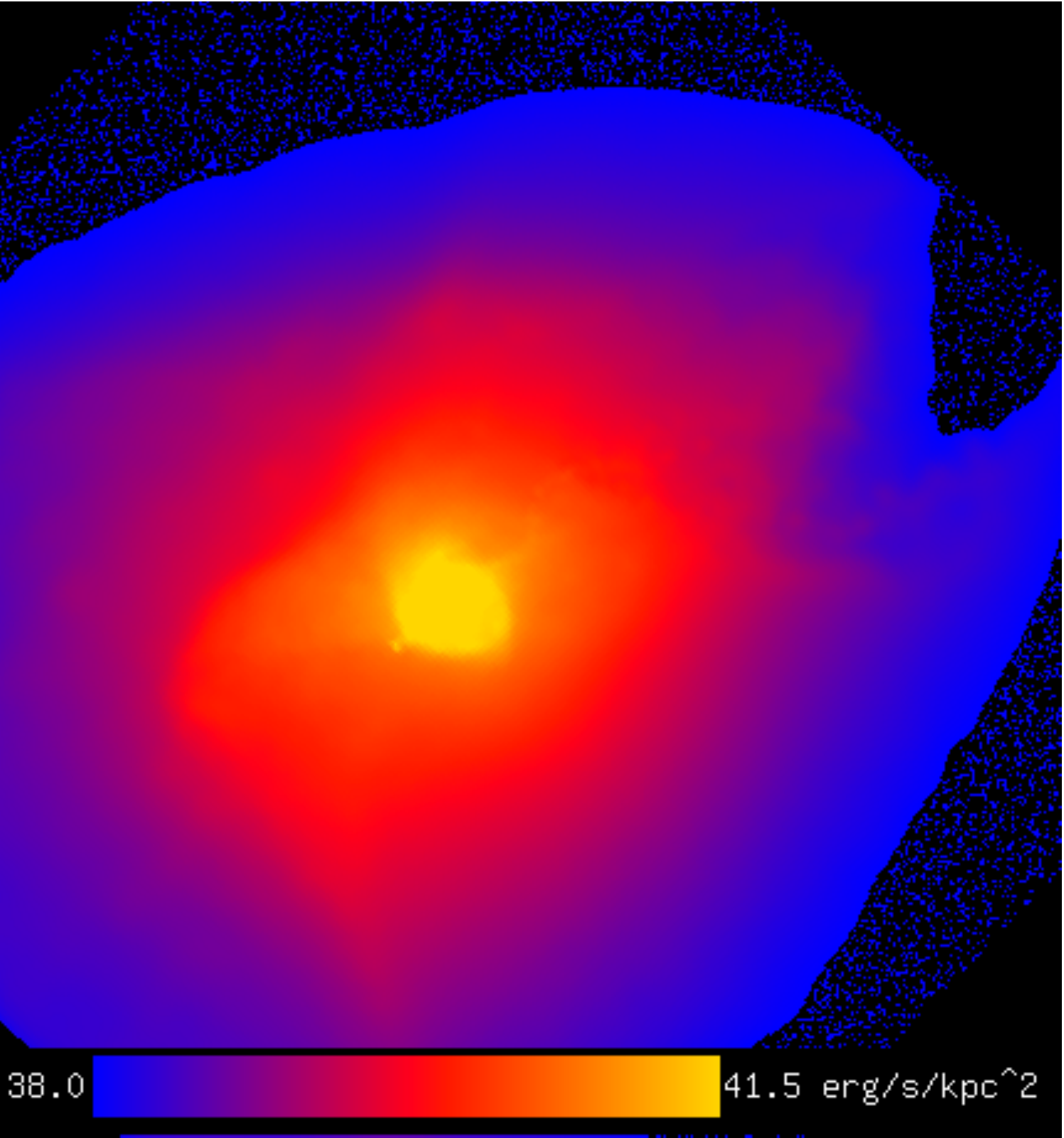}}\hspace{0.05cm}
\caption{Surface brightness (middle row, left panel) and gas temperature distribution (middle row, right panel) obtained from simulations of an off-axis, 3:1 binary cluster merger, caught at 0.4 Gyr after pericentric passage. At this moment, our observations represent a good match to the simulated data. Overlaid are Chandra X-ray surface brightness contours of \ab295. The top and bottom row of panels show the surface brightness from simulated data corresponding to different instances of time measured relative to $t_0$, the time when the secondary traverses a circular region of radius $\rm R_{200}$ centered on the primary: for top row - 0 Gyr, 0.3 Gyr, 0.5 Gyr (moment of pericentric passage), 0.7 Gyr; for middle row - 0.9 Gyr; for bottom row - 1.1 Gyr, 1.5 Gyr (moment of apocentric passage), 1.9 Gyr, 2.5 Gyr (moment of second pericentric passage). The figure in each panel was scaled and rotated such as to match the observed data. All figures were adapted from  \protect\cite{Poole2006}.}
\label{fig:poole_simulations}
\end{figure*}

Figure \ref{fig:poole_simulations}, which is adapted from the results of \cite{Poole2006}, shows in the central row of two panels the surface brightness (left) and temperature (right) profiles corresponding to one of their hydrodynamical simulations of binary mergers. From all the available results, obtained for a large range of merger parameters, this simulation has been found to agree the best, in a qualitative way, with our observations. For this particular simulation, the mass ratio between primary and secondary cluster is 3:1, with  the primary having a mass of $\rm 10^{15} \Msol$.  It is an off-axis merger, caught at a time during the secondary's first travel between pericentre and apocentre. The secondary moves from SE to NW  (in the rotated image). The top and bottom row of panels show the surface brightness distribution for the same merger, but at different stages during the merging process: the first pericentric passage and several epochs before this (top panels) and  the apocentric passage and the 
second closest encounter of the two clusters (bottom panels). 

Looking firstly at the surface brightness map (left panel in the central row of panels) we see a good agreement between the morphology of \ab295 as shown by the X-ray surface brightness contours and the surface brightness obtained from simulations. We see the same elongated central surface brightness which joins the two subclusters. The plume observed to the south of the primary cluster matches very well the position of a surface brightness edge seen in the simulated image. At larger radii, the cluster has a relaxed morphology, with no obvious surface brightness features. Also, there is a steepening of surface brightness gradient in front of the secondary cluster. A significant difference between the properties of \ab295 and the simulated results is the absence of a cool core in the primary of \ab295, while Poole's simulations, like most cluster merger simulations, follow the merger of clusters in which both systems have a central cool core, visible as a significant surface brightness excess and temperature 
drop at 
the cluster center.

Our observed temperature map lacks the necessary spatial resolution to make a detailed comparison with simulations. However, the cool region associated with the secondary's core (regions 1 and 2) and the hot region to the SE of the main cluster (region 32) are the main features in the observed data for which we found a match in the simulated temperature map. 

The presence of a cold front in the secondary cluster, as seen in our data (see Section \ref{subsec:CF}), is in good agreement with predictions from these simulations. In most binary cluster, off-axis merger simulations a cold front starts forming in front of the secondary shortly before the first pericentric passage and it survives for a long time, typically until the second pericentric passage after which it mixes with the gas in the primary cluster. During its lifetime, the cold front suffers changes in its morphology and strength depending on the merger state and parameters.  Immediately after first pericentric passage, the sudden release of ram pressure exerted on the gas in the secondary by the primary's gas triggers the gravitational slingshot process \citep{Hallman2004}. The gas will overshoot the dark mater component, will expand adiabatically and cool further, making the cold front increase its radius of curvature to values which can reach several kiloparsecs and enhance its temperature jump.

While there is general agreement with the simulations presented in Figure \ref{fig:poole_simulations}, there is a significant difference between the morphology of the cold front in \ab295 and that seen in the simulations. Such a difference is not surprising since the simulations presented in the Figure are not tailored to match \ab295 and the  morphology of cold fronts is dependent on the merger parameters. However, the sole detection of a cold front in the secondary and its properties can be used to put some constraints on the cluster dynamics. 

One feature that matches very well the simulated results is the plume seen to the south of the primary cluster. It can be seen as an edge in surface brightness as well as a drop in temperature.
This feature is seen in most simulations of binary cluster mergers and is interpreted as gas displaced from the primary core as a result of the interaction between the two clusters, close to the time of the first pericentric passage. At this time, the gas in the primary expands as it goes through the slingshot process. The expansion of the gas will be confined to one side by the shock in front of the secondary, hence the linearity of the plume.   

As most simulations follow the merging of two cool core clusters, there is already a reservoir of cool gas in the primary which might represent the source for the gas in the plume. This raises the question whether a cool core is necessary for the formation of this feature. \cite{Ascasibar2006} have shown that this feature can form even if the primary has a flat distribution of central temperature profile. This is also the case of \ab295, in which no cool gas has been found at the core of the primary. However, in addition to the plume, their results predict that a cold front, of slingshot origin, should be visible in the primary, a feature which has not been detected in our data. 

Although this feature has been seen for a wide range of impact parameters and mass ratios for off-axis simulations, observationally, there is a very limited number of plumes detected in merging clusters. The best example is Abell 2146 \citep{Russell2010}, which has many similarities with \ab295.  Very large mass ratios, impact parameters or angle of merger axis may be amongst the reasons for a lack of observation of plumes in most merging clusters. 

Since a large inclination angle of the merger axis can easily hide the plume due to projection effects, the detection of this feature in \ab295 constrains the inclination of the merger axis with respect to the plane of the sky  to relatively small values. Moreover, the presence of the plume rules out very large impact parameters for the merger, since this would not be able to significantly disturb the core of the primary.

\subsection{Shocks in \ab295?}
\label{subsec:shocks}

Two important features predicted by the \cite{Poole2006} simulations are the two outwardly propagating shocks which form shortly before first pericentric passage: a bow shock in front of the secondary and a reverse shock in the vicinity of the primary. The two shocks are clearly seen as the hottest regions in the temperature map obtained from simulations and presented in Figure \ref{fig:poole_simulations}. In \ab295 we see little to no indication of a bow shock and some weak evidence in favor of the presence of a reverse shock.

Regarding the bow shock, we have shown in Section \ref{subsec:CF} that besides the surface brightness discontinuity associated with the cold front, there is another break, at a distance of 246 kpc from the cold front, where the density drops by a factor of $2.13 \pm 0.87$ (see also Table \ref{table:jump_prop}). Although the temperature profile shows a drop in temperature across this discontinuity ($1.69_{-0.63}^{+1.15}$), it is not significant enough (1 sigma) to confirm the presence of a shock. If this density and temperature jump correspond to a shock, then its Mach number, as estimated from the amplitude of density and temperature jump, is $1.85 \pm 0.81 $ and $1.49_{-0.67}^{+1.24}$, respectively. 

Additional evidence supporting the presence of a bow shock comes from radio observations of \ab295 \citep{Zheng2018}, which show diffuse radio emission in the vicinity of the detected surface brightness discontinuity (compare the position of the radio halo in Figure \ref{fig:AS0295_image}, right panel, with Figure \ref{fig:unsharp-masked}). From the properties of this emission, interpreted as a radio relic, the authors derived a Mach number for the bow shock of $2.04$. Their results are consistent with our finding that the bow shock in \ab295 is in the weak shock regime. 

With respect to the reverse shock, we found some weak evidence in favor of the existence of a shock (see Section \ref{subsec:SE_discontinuity}). The Mach numbers estimated from the temperature and density jump ($1.69_{-0.43}^{+0.58}$ and $1.24_{-0.02}^{+0.02}$, respectively) agree within the errors and correspond, like in the case of the bow shock, to a weak shock. \cite{Zheng2018} claim the presence of another radio relic at the position of this shock. However, the characterization of this diffuse emission as a radio relic is hindered by the difficulty in the removal of the contribution of a complex radio point source embedded in the diffuse emission.

 \begin{table*}
    \caption{Proprieties of three detected surface brightness jumps (Section \ref{sec:as0295}). Columns contain information about the region where the jump was detected (col 1), the proposed nature of the feature (col 2), ratio between temperature (col 3) and density (col 4) across the jump and, in case the feature is characterized as a shock, the estimated Mach number, calculated as weighted mean from Mach numbers estimated from temperature and density jumps. }
    \label{table:jump_prop}
    \begin{center}
	  \begin{tabular}{ccccc}
	  \hline
	  \\
	  Region & Feature type  & $\rm \frac{kT_2}{kT_1}$ & $\rm \frac{n_{e2}}{n_{e1}}$ &Mach number\\
	  \\
	  \cline{1-5}\\
		      NW & cold front & $0.51_{-0.13}^{+0.18}$  & $2.19_{-0.29}^{+0.29}$ & --  \vspace{0.2cm} \\
		      NW & shock? & $1.69_{-0.63}^{+1.15}$ & $2.13_{-0.87}^{+0.87}$& $1.74_{-0.74}^{+1.02}$   \vspace{0.2cm}\\
		      SE & shock? &  $1.89_{-0.45}^{+0.61}$ & $1.35 _{-0.03}^{+0.03}$& $1.24_{-0.22}^{+0.30}$ \vspace{0.2cm}\\
	  \end{tabular}
    \end{center}
\end{table*}

\subsection{Possible merger scenario}
\label{subsec:merger_scenario}
						  
Based on the observed X-ray data and the comparison of the two most significant features detected in the surface brightness distribution (see Sections \ref{subsec:CF} and \ref{subsec:plume}) as well as the comparison of the global proprieties of \ab295 with simulation results (see Section \ref{subsec:comparison_sim}), we propose an off-axis, binary major merger scenario for \ab295 in which the secondary is in its way to the apocentre, after having its first closest encounter with the core of primary.  

More constraints on this merging scenario can be obtained by adding information from the strong lensing analysis of \ab295 by \cite{Cibirka2018}. Their results show two well separated mass peaks, associated with the individual subclusters (see Figure \ref{fig:AS0295_image}, right panel), thus clearly confirming the binary merger scenario. Comparison of the strong lensing mass map with the optical image (Figure \ref{fig:AS0295_image}, central panel) shows that the distribution of cluster galaxies traces the mass distribution peaks.

Numerical simulations (\cite{Ricker2001}; \cite{Poole2006}; \cite{Ascasibar2006}; \cite{ZuHone2011}) show that in most binary mergers,  an offset between the position of the gas and dark matter in the primary and/or secondary is created at some particular stages during the merger. While around the moment of core passage, the high ram pressure created by the primary's gas on the secondary leads to an offset between the secondary's gas and dark matter, with the dark matter component leading the gas, immediately after core passage, the ram-pressure drops quickly. As a result, the  gas goes through the gravitational slingshot process \citep{Hallman2004} and starts overtaking the dark matter. The exact moment when the gas reaches the dark matter depends on the merging parameters such as mass ratio, impact parameter and initial relative velocities of the two subclusters. 

Similarly to the secondary, the gas in the core of the primary feels the effect of the high ram pressure generated at core passage which leads to an offset between the gas and dark matter. 

These gas-dark matter offsets in the primary and secondary can be used to put constraints on the merging scenario. For \ab295 we found that the position of the secondary's mass peak (coincident with the position of the BCG) matches that of the peak of X-ray emission within the core of secondary. 
Although there is no X-ray peak associated with the primary cluster, we see that the mass peak is well separated from the high surface brightness region corresponding to the bulk of the gas in the core. Such a large offset is not seen in merging simulations of cool core clusters, where the compact cool core tends to follow more closely the DM component. However, when the primary has a flat central surface brightness distribution, the gas is easily separated from the dark matter \citep{Ascasibar2006,Mastropietro2008}. Based on the above mentioned observations, we propose a merger scenario in which the secondary is caught at a time shortly after the pericentric passage.

\subsection{Limits on DM self-interaction cross-section}
Cluster mergers, especially those for which there is a clear offset between the gas and the DM  component, are ideal sources for the study of DM properties. In these systems, the existence of this offset implies different scattering depths for the two components, and therefore a constraint on the DM can be obtained by comparison of its optical depth with that of the gas \citep{Markevitch2004}. In AS0295, we see a displacement of the DM from the bulk of the gas in the primary, while for the secondary, the positions of the DM and gas peaks are identical.  In the following we apply the usual formalism to the DM/gas offset of the primary component.

Following \cite{Markevitch2004}, we have estimated the DM self-interaction cross-section under the condition that the scattering depth of dark matter ($\tau$) cannot be much greater than 1. The scattering depth is given by\\
\begin{equation*}
\tau=\dfrac{\sigma}{m}\Sigma,
\end{equation*}
where $\sigma$ is the DM self-interaction cross-section, $m$ is the DM particle mass and $\Sigma$ is the DM surface density. Under the assumption of spherical symmetry, the surface mass density along the collision direction equals the surface mass density along the line of sight. We estimated the surface mass density from lensing maps of \cite{Cibirka2018} using a circular region, with radius of 150 kpc, centered on the primary component's centroid, although the numerical value is not very sensitive ($<$10\%) to the precise location. 

Based on the estimated surface mass density of $\rm \sim 0.35 ~ g cm^{-2}$, we obtain an estimate for the self-interaction cross-section of $\sigma /m <3 ~\rm cm^2 g^{-1}$.   The derived upper limit is comparable to other cross-section estimates (ranging between $3-7 ~\rm cm^2 g^{-1}$) based on the scattering depth method and reported in different studies of cluster mergers (  Bullet cluster - \cite{Markevitch2004}; MACS J0025.4-1222 - \cite{Bradavc2008}; Abell 2744 - \cite{Merten2011};  DLSCL J0916.2+2951 - \cite{Dawson2012} ).

\section{Conclusions}
\label{sec:conclusions}

We have presented \Chandra observations of \ab295 cluster, a hot ($9.5~ \rm keV$) cluster at redshift $0.3$, going through a merging process. 

The binary nature of the merger is suggested by the morphology of the cluster, with an evident surface brightness peak associated with the secondary and well separated from the bulk of the primary's emission. The distribution of galaxies in the optical image and the presence of two separate mass peaks visible in the strong lensing map from \cite{Cibirka2018} reinforce the binary merger scenario. The orientation of the merger axis in the SE-NW direction is indicated by the position of the secondary relative to the primary cluster and the elongation of the surface brightness distribution in the SE-NW direction. 

In the temperature map, we found gas as cool as 6 keV associated with the core of the secondary cluster, while the primary has a central gas temperature comparable to the cluster mean, which is typical for a non-cool core cluster. We found a small region of hot gas ($\sim20\, \rm keV$) in the primary, which is offset from the central bulk of the gas toward a low surface brightness region. We interpret the hot region, together with the surface brightness discontinuity detected in its close vicinity, as a reverse shock created when the clusters reached their closest approach. For this shock we  derive a Mach number of $1.24_{-0.22}^{+0.30}$. We also found weak evidence for the existence of another shock leading the secondary, for which we derived a Mach number of $1.74_{-0.74}^{+1.02}$.

In addition to these probable shocks, we detect two other merger signatures: a cold front ahead of the secondary and a plume of cool gas emerging from the primary. While cold fronts are a commonly detected feature in merging cluster observations, the plume, which is a feature expected in off-axis, binary cluster simulations has been reported in a more limited number of cases (the best example is Abell 2142 \cite{Russell2012} with which \ab295 has many similarities). 

Good agreement was found when comparing the surface brightness and temperature distribution of \ab295 with simulations of binary mergers. Based on our data and the simulations, we propose a merging scenario in which the cluster is an off-axis merger, with the secondary on its way to apocentre or close to reaching it. 

In the primary cluster component there is a significant spatial offset between the peaks of the gas (from the X-ray image) and the dark matter (from a strong-lensing surface-mass image) from which we obtain an estimate for the self-interaction cross-section of $\sigma /m <3 ~\rm cm^2 g^{-1}$.

\section*{Acknowledgments}
This research has made use of data obtained from the Chandra Data Archive and the Chandra Source Catalog, and software provided by the Chandra X-ray Center (CXC) in the application packages CIAO, ChIPS, and Sherpa.  JPH acknowledges support from the National Science Foundation through Astronomy and Astrophysics Research Program award number 1615657.  This study used
observations made with the NASA/ESA Hubble Space Telescope, and obtained from the Hubble Legacy Archive, which is a collaboration between the Space Telescope Science Institute (STScI/NASA), the Space Telescope European Coordinating Facility (ST-ECF/ESA) and the Canadian Astronomy Data Centre (CADC/NRC/CSA).
\bibliography{AS0295}

\clearpage

\end{document}